\documentclass[]{spie}  
\usepackage[T1]{fontenc} 

\usepackage{graphicx}
\usepackage{transparent}
\usepackage{xkeyval,xcolor}
\makeatletter
\newlength{\sfp@hseplen}\newlength{\sfp@vseplen}
\define@cmdkey{subfigpos}[sfp@]{pos}[ul]{}
\define@cmdkey{subfigpos}[sfp@]{font}[\small]{}
\define@cmdkey{subfigpos}[sfp@]{vsep}[0.75\baselineskip]{\setlength{\sfp@vseplen}{\sfp@vsep}}
\define@cmdkey{subfigpos}[sfp@]{hsep}[3.5pt]{\setlength{\sfp@hseplen}{\sfp@hsep}}
\newcommand{\subfigimg}[4][,]{%
        \setkeys{Gin,subfigpos}{pos,font,vsep,hsep,#1}
        \setbox1=\hbox{\includegraphics{#4}}
        \ifnum\pdfstrcmp{\sfp@pos}{ul}=0
                \leavevmode\rlap{\usebox1}
                \rlap{\hspace*{\sfp@hsep}\raisebox{\dimexpr\ht1-\sfp@vsep}{\transparent{#3}{\setlength{\fboxsep}{1pt}\colorbox{white}{%
\transparent{1}\sfp@font{#2}}}%
}}
                \phantom{\usebox1}
        \else\ifnum\pdfstrcmp{\sfp@pos}{ur}=0
                \leavevmode\usebox1
                \llap{\raisebox{\dimexpr\ht1-\sfp@vsep}{\sfp@font{#2}}\hspace*{\sfp@hsep}}
        \else\ifnum\pdfstrcmp{\sfp@pos}{lr}=0
                \leavevmode\usebox1
                \llap{\raisebox{\sfp@vsep}{\sfp@font{#2}}\hspace*{\sfp@hsep}}
        \else
                \leavevmode\rlap{\usebox1}
                \rlap{\hspace*{\sfp@hseplen}\raisebox{\sfp@vsep}{\sfp@font{#2}}}
                \phantom{\usebox1}
        \fi\fi\fi
}
\newcommand{\fontfig}[1]{\tiny$\!\!$\color{#1}\textbf}
\newcommand{\AspectRatio}[1]{\dimexpr 1pt * \wd#1 / \ht#1 \relax} 

\usepackage{amsmath} 
\usepackage{array}
\usepackage{adjustbox}
\usepackage{multirow}
\usepackage{makecell}
\newcolumntype{C}[1]{>{\centering\arraybackslash}p{#1}} 

\newcommand{\subfigref}[1]{(#1)} 
\newcommand{\fig}[1]{Fig.~\ref{#1}\xspace}
\newcommand{\figs}[2]{Figs.~\ref{#1} and~\ref{#2}\xspace} 
\newcommand{\figfull}[1]{Figure~\ref{#1}\xspace} 
\newcommand{\subfig}[2]{Fig.~\ref{#1}\subfigref{#2}\xspace}
\newcommand{\subfigfull}[2]{Figure~\ref{#1}\subfigref{#2}\xspace} 
\newcommand{\subfigs}[2]{Figs.~\ref{#1}\subfigref{#2}\xspace}
\newcommand{\subfigsfull}[2]{Figures~\ref{#1}\subfigref{#2}\xspace} 

\newcommand{\tab}[1]{Table~\ref{#1}\xspace}

\newcommand{\refsec}[1]{Section~\ref{#1}\xspace} 
\newcommand{\refapp}[1]{Appendix~\ref{#1}\xspace} 

\usepackage{algorithm}
\usepackage[noend]{algpseudocode}
\newcommand{\commentalgo}[1]{\Comment{{\tiny #1}}} 

\usepackage{mathtools}
\usepackage{mathrsfs}
\usepackage{nicefrac}
\usepackage{stmaryrd} 
\SetSymbolFont{stmry}{bold}{U}{stmry}{m}{n} 

\newcommand{\Tag}[1]{\text{#1}}   

\newcommand{\Lens}[1]{${\text{\textbf{L}}{\boldsymbol{#1}}}$} 
\newcommand{\FM}[1]{${\text{\textbf{FM}}_{\boldsymbol{#1}}}$} 

\newcommand{\V}[1]{{\boldsymbol{#1}}}                 
\newcommand{\Inv}{^{-1}}                              

\DeclarePairedDelimiterX{\paren}[1]{(}{)}{#1}
\newcommand{\Paren}[1]{\paren*{#1}}
\let\brace=\undefined 
\DeclarePairedDelimiterX{\brace}[1]{\{}{\}}{#1}
\newcommand{\Brace}[1]{\brace*{#1}}
\let\brack=\undefined 
\DeclarePairedDelimiterX{\brack}[1]{[}{]}{#1}
\newcommand{\Brack}[1]{\brack*{#1}}
\DeclarePairedDelimiterX{\bbrack}[1]{\llbracket}{\rrbracket}{#1}
\newcommand{\Bbrack}[1]{\bbrack*{#1}}
\DeclarePairedDelimiterX{\abs}[1]{\rvert}{\lvert}{#1}     

\DeclarePairedDelimiterX{\norm}[1]{\lVert}{\rVert}{#1}    
\newcommand{\Norm}[1]{\norm*{#1}}

\DeclarePairedDelimiterX{\avg}[1]{\langle}{\rangle}{#1}   

\DeclarePairedDelimiterX{\ceil}[1]{\lceil}{\rceil}{#1}     

\DeclarePairedDelimiterX{\floor}[1]{\lfloor}{\rfloor}{#1}  

\newcommand{\Vx}{{\V{x}}}                   
\newcommand{\Vxp}{{\V{x^{\prime}}}}         

\newcommand{\AvgT}[1]{\avg*{#1}}          
\newcommand{\pha}{\varphi}                  
\newcommand{\SF}[1]{D_{#1}}         
\newcommand{\rFried}{r_{0}}             
\newcommand{\seeing}{\epsilon}             

\newcommand{\na}{n_\Tag{a}}	
\newcommand{\nm}{n_\Tag{m}}	
\newcommand{\IFa}{\pha_{a}}	
\newcommand{\phaBG}{\pha_{\Tag{bg}}}	
\newcommand{\pham}{\pha_{\Tag{m}}}   
\newcommand{\pix}{\theta} 	
\newcommand{\Vpix}{\V{\pix}}	
\newcommand{\IFrad}{\psi}	
\newcommand{\IFrada}{\IFrad_{a}} 
\newcommand{\IFradnot}{\IFrad_{0}} 
\newcommand{\mask}{\mathcal{M}}	

\newcommand{\stroke}{\includegraphics[height={\f@size pt*2/3}]{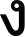}}
\newcommand{\slope}{\tilde{\alpha}}
\newcommand{\offset}{\tilde{o}}
\DeclarePairedDelimiterX{\clip}[1]{\lfloor}{\rceil}{#1}
\newcommand{\Clip}[1]{\clip*{#1}}
\newcommand{\ClipVal}[3]{\Clip{#1}_{#2}^{#3}}

\newcommand{\etal}{\textit{et al.}\xspace}
\newcommand{\vs}{v.s.\xspace}

\usepackage{siunitx}
\newcommand{\inch}[1]{\SI{#1}{''}} 

\usepackage[colorlinks=true, allcolors=blue]{hyperref}
\usepackage{xspace}

\title{Adaptive Optics system of the Evanescent Wave Coronagraph (EvWaCo) -- optimised phase plate and DM characterisation}

\author[a,b]{Anthony Berdeu}
\author[a]{Sitthichat Sukpholtham}
\author[a]{Puttiwat Kongkaew}
\author[a]{Adithep Kawinkij}
\author[a]{Matthew Ridsdill-Smith}
\author[c]{Michel Tallon}
\author[c]{Éric Thiébaut}
\author[c]{Maud Langlois}
\author[a]{Mary Angelie Alagao}

\affil[a]{National Astronomical Research Institute of Thailand, Center for Optics and Photonics, 260 Moo 4, T. Donkaew, A. Maerim, Chiang Mai 50180, Thailand}
\affil[b]{Department of Physics, Faculty of Science, Chulalongkorn University, 254 Phayathai Road, Pathumwan, Bangkok 10330, Thailand}
\affil[c]{Univ Lyon, Univ Lyon1, Ens de Lyon, Centre de Recherche Astrophysique de Lyon, UMR 5574, F-69230, Saint-Genis-Laval, France}

\authorinfo{
Send correspondence to Anthony Berdeu: \href{mailto:anthony@narit.or.th}{anthony@narit.or.th}
}

\begin{document} 
\maketitle

\begin{abstract}
The Evanescent Wave Coronagraph (EvWaCo) is an achromatic coronagraph mask with adjustable size over the spectral domain [\SI{600}{\nano\meter}, \SI{900}{\nano\meter}] that will be installed at the Thai National Observatory. We present in this work the development of a bench to characterise its Extreme Adaptive Optics system (XAO) comprising a DM192 ALPAO deformable mirror (DM) and a 15x15 Shack-Hartmann wavefront sensor (SH-WFS). In this bench, the turbulence is simulated using a rotating phase plate in a pupil plane. In general, such components are designed using a randomly generated phase screen. Such single realisation does not necessarily provide the wanted structure function. We present a solution to design the printed pattern to ensure that the beam sees a strict and controlled Kolmogorov statistics with the correct 2D structure function. This is essential to control the experimental conditions in order to compare the bench results with the numerical simulations and predictions. This bench is further used to deeply characterise the full 27 mm pupil of the ALPAO DM using a $54\times54$ ALPAO SH-WFS. We measure the average shape of its influence functions as well as the influence function of each single actuator to study their dispersion. We study the linearity of the actuator amplitude with the command as well as the linearity of the influence function profile. We also study the actuator offsets as well as the membrane shape at 0-command. This knowledge is critical to get a forward model of the DM for the XAO control loop.
\end{abstract}


\section{INTRODUCTION}

The Evanescent Wave Coronagraph (EvWaCo\cite{Buisset:17_EvWaCo}) is an on-going project at the National Astronomical Research Institute of Thailand (NARIT). EvWaCo is a coronagraph that uses the frustration of the total internal reflection (FTIR\cite{Zhu:86_FTIR}) between a prism and a lens put in contact. The star light, passes through the contact area while the light of the stellar surrounding is internally reflected in the prism\cite{Alagao:21_EvWaCo_exp_contrast}. The FTIR being wavelength dependent, the mask scales with the wavelength and thus presents almost achromatic contrast performances over the spectral domain~$\Brack{\SI{600}{\nano\meter},\SI{900}{\nano\meter}}$. In addition, by controlling the pressure between the two pieces of glass, the size of the mask can be finely adjusted.

An on-sky demonstrator is currently under development at NARIT to be installed on the \SI{2.4}{\meter} Thai National Telescope, using an elliptical unobstructed pupil of $1.17\times\SI{0.83}{\meter\squared}$. This implies the development of a dedicated adaptive optics (AO) system. A preliminary design of this prototype and its AO system was presented in a previous work\cite{Buisset:18_EvWaCo_spec}. It was based on a deformable mirror (DM) with $\na=192$ actuators from ALPAO (the DM192), and a draft for the Shack-Hartmann (SH) wavefront sensor (WFS) of EvWaCo. The design is now finalised and most of the components have been procured.

\begin{figure}[ht!]
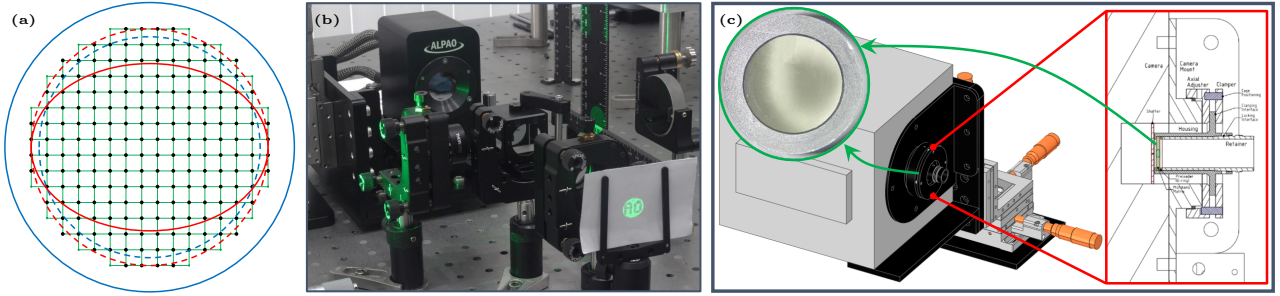
 
        \centering
        \newcommand{\LineRatio}{0.975}
        
        \newcommand{\PathFig}{figures_DM192_}
        \newcommand{\FigOne}{\PathFig figa.pdf}
        \newcommand{\FigTwo}{\PathFig figb.pdf}
        \newcommand{\FigThree}{\PathFig figc.pdf}
        
        \newcommand{\subfigColor}{black}        
        
        \sbox1{\includegraphics{\FigOne}}               
        \sbox2{\includegraphics{\FigTwo}}               
        \sbox3{\includegraphics{\FigThree}}               
        \newcommand{\ColumnWidth}[1]
                {\dimexpr \LineRatio \linewidth * \AspectRatio{#1} / (\AspectRatio{1} + \AspectRatio{2} + \AspectRatio{3}) \relax
                }
        \newcommand{\ColumnGap}{\hspace {\dimexpr \linewidth /4 - \LineRatio\linewidth /4 }}

        \begin{tabular}{
                @{\ColumnGap}
                C{\ColumnWidth{1}}
                @{\ColumnGap}
                C{\ColumnWidth{2}}
                @{\ColumnGap}
                C{\ColumnWidth{3}}
                @{\ColumnGap}
                }
                \subfigimg[width=\linewidth,pos=ul,font=\fontfig{\subfigColor}]{$\,$(a)}{0.0}{\FigOne} &
                \subfigimg[width=\linewidth,pos=ul,font=\fontfig{\subfigColor}]{$\,$(b)}{0.0}{\FigTwo} &
                \subfigimg[width=\linewidth,pos=ul,font=\fontfig{\subfigColor}]{$\,$(c)}{0.0}{\FigThree}
        \end{tabular}        
        
        \caption{\label{fig:DM192} Overview of the EvWaCo AO system: deformable mirror and Shack-Hartmann wavefront sensor. (a) Geometry of the EvWaCo WFS on the DM192: $15\times11$ sub-apertures, $16\times12$ actuators, no obstruction. Red ellipse: $1.17\times\SI{0.83}{\meter\squared}$ EvWaCo pupil. Red dashed circle: \SI{22.5}{\milli\meter} circular pupil. The original \SI{21}{\milli\meter} circular stop of the DM (blue dashed circle) was replaced with a \SI{27.5}{\milli\meter} stop (blue circle). The black dots are the actuator positions, placed on a Cartesian grid of pitch \SI{1.5}{\milli\meter}. The green squares represent the lenslets of the WFS in the Fried geometry. (b) Picture of the DM192 in the NARIT cleanroom on its mount designed and manufactured at NARIT. (c) Schematic of the EvWaCo WFS: the mount for the Nüvü camera and the adaptor and clamping system (framed in red) to insert the lenslet array (circled in green) inside the camera flange have been designed and manufactured at NARIT.}
\end{figure}

\subfigfull{fig:DM192}{a} introduces the geometry of the AO system in the DM pupil frame and the main features of the EvWaCo AO system are summed up in \tab{tab:AO_EvWaCo}. On our demand, ALPAO kindly changed the original \SI{21}{\milli\meter} circular stop of the DM192 by a larger stop of \SI{27.5}{\milli\meter} in diameter, see \subfig{fig:DM192}{b}. Doing so, the AO system of EvWaCo uses the full diameter of \SI{22.5}{\milli\meter} delimited by the extreme most actuators across the DM diameter (red ellipse on the figure), spanning over $16\times12$ actuators with a pitch of \SI{1.5}{\milli\meter}. In the Fried geometry\cite{Fried:77, Southwell:80}, this corresponds to $15\times11$ sub-apertures with the actuators placed on their corners (green squares on the figure).

\begin{table}[!ht] 
	\centering
	\caption{\centering\label{tab:AO_EvWaCo} Main features of the EvWaCo AO system and its SH-WFS. The first sub-table lists some important features of the Nüvü~$128^\Tag{AO}$ EMCCD detector as specified by Nüvü Cam$\bar{\text{e}}$ras. The second sub-table lists the parameters of the EvWaCo SH-WFS. The third sub-table lists the parameters of the procured lenslet array as specified by Smart MicroOptical Solution company.}
	\begin{tabular}{|l|l|}
	\hline
	Number of pixels & $128\times128$
	\\ \hline
	Pixel size & $24\times\SI{24}{\micro\meter}$
	\\ \hline
	Chip size & $3.072\times\SI{3.072}{\milli\meter\squared}$
	\\ \hline
	Digitisation & \SI{16}{bits}
	\\ \hline
	Maximal frame rate & $\SI{1004}{\hertz}$
	\\ \hline
	$\text{e}^{-}$-multiplying gain & $1\leftrightarrow5000$
	\\ \hline
	Dark current at \SI{-60}{\celsius} & $\SI{0.02}{\text{e}^{-}/\text{pixel}/\text{s}}$
	\\ \hline
	Clock-induced charges & $\SI{0.005}{\text{e}^{-}/\text{pixel}/\text{frame}}$ at $\text{gain}=1000$
	\\ \hline
	\hline
	Number of sub-apertures & $15\times11$, \subfig{fig:DM192}{a}
	\\ \hline
	Average wavelength & $\SI{550}{\nano\meter}$
	\\ \hline
	Number of pixels per sub-aperture & $\SI{8}{\text{pixels}}\times\SI{8}{\text{pixels}}$
	\\ \hline
	Size of a lenslet & $192\times\SI{192}{\micro\meter}$
	\\ \hline
	Size of the pupil & $2.88\times\SI{2.04}{\milli\meter}$
	\\ \hline
	Lenslet focal length & $\SI{15}{\milli\meter}$
	\\ \hline
	$\lambda/d$ & $\SI{1.5}{''}$
	\\ \hline
	Pixel scale & $\SI{0.8}{''}$
	\\ \hline
	Field of view & $\SI{6.4}{''}\times\SI{6.4}{''}$
	\\ \hline
	\hline
	Back distance array - focal point & $\SI{15}{\milli\meter}$
	\\ \hline
	Size of square lenslet (\SI{100}{\percent} fill factor) & $192\pm\SI{0.1}{\micro\meter}$
	\\ \hline
	Wavefront quality & $<\lambda/10$
	\\ \hline
	Wavelength range & $400\leftrightarrow\SI{800}{\nano\meter}$
	\\ \hline
	\end{tabular}
\end{table}

The camera of the EvWaCo WFS is the Nüvü~$128^\Tag{AO}$ from Nüvü Cam$\bar{\text{e}}$ras\footnote{\url{https://www.nuvucameras.com}}. It was chosen for its electron-multiplying charge-coupled device (EMCCD) detector with a high frame rate and very low noise, see \tab{tab:AO_EvWaCo}. To avoid any relay lens system and thus keep the AO system compact, the lenslet array of the SH-WFS is designed to be introduced directly in front of the Nüvü sensor. This implies a custom-designed lenslet array that was manufactured by Smart MicroOptical Solution company\cite{Bahr:15_lenslet}.

Finally, dedicated mounts have been designed and manufactured at NARIT for the DM192 and the Nüvü camera, see \subfigs{fig:DM192}{b,c}. An adaptor and a fine tuning system have also been designed and manufactured at NARIT to place, position, orientate and clamp the lenslet array inside the flange of the camera, see \subfig{fig:DM192}{c}.

Before assembling the AO system inside the EvWaCo prototype, its different components must be tested and characterised. To do so, a dedicated AO bench has been designed and built in the  NARIT cleanroom. In addition to the component characterisation, this bench is also used for the AO loop algorithm development from the coding of the hardware control to the assessment of the AO loop performances\cite{Berdeu:22_AO_loop} and predict on-sky contrast\cite{Ridsdill:22}.

This paper focuses on this AO bench design and characterisation. First, \refsec{sec:AO_bench} gives an overview of the bench. Then,  \refsec{sec:PP} focuses on the design and the characterisation of a phase plate (PP) from Lexitek\cite{Ebstein:02_Lexitek} used to emulate the atmospheric turbulence in the bench. Finally, \refsec{sec:DM} presents the detailed analysis performed on the DM192 from ALPAO\cite{LeBouquin:18_charac_ALPAO}\footnote{\url{http://www.alpao.com}}.

\section{AO BENCH DESCRIPTION}
\label{sec:AO_bench}

\begin{figure}[ht!] 
        \centering
        \includegraphics[width=\linewidth]{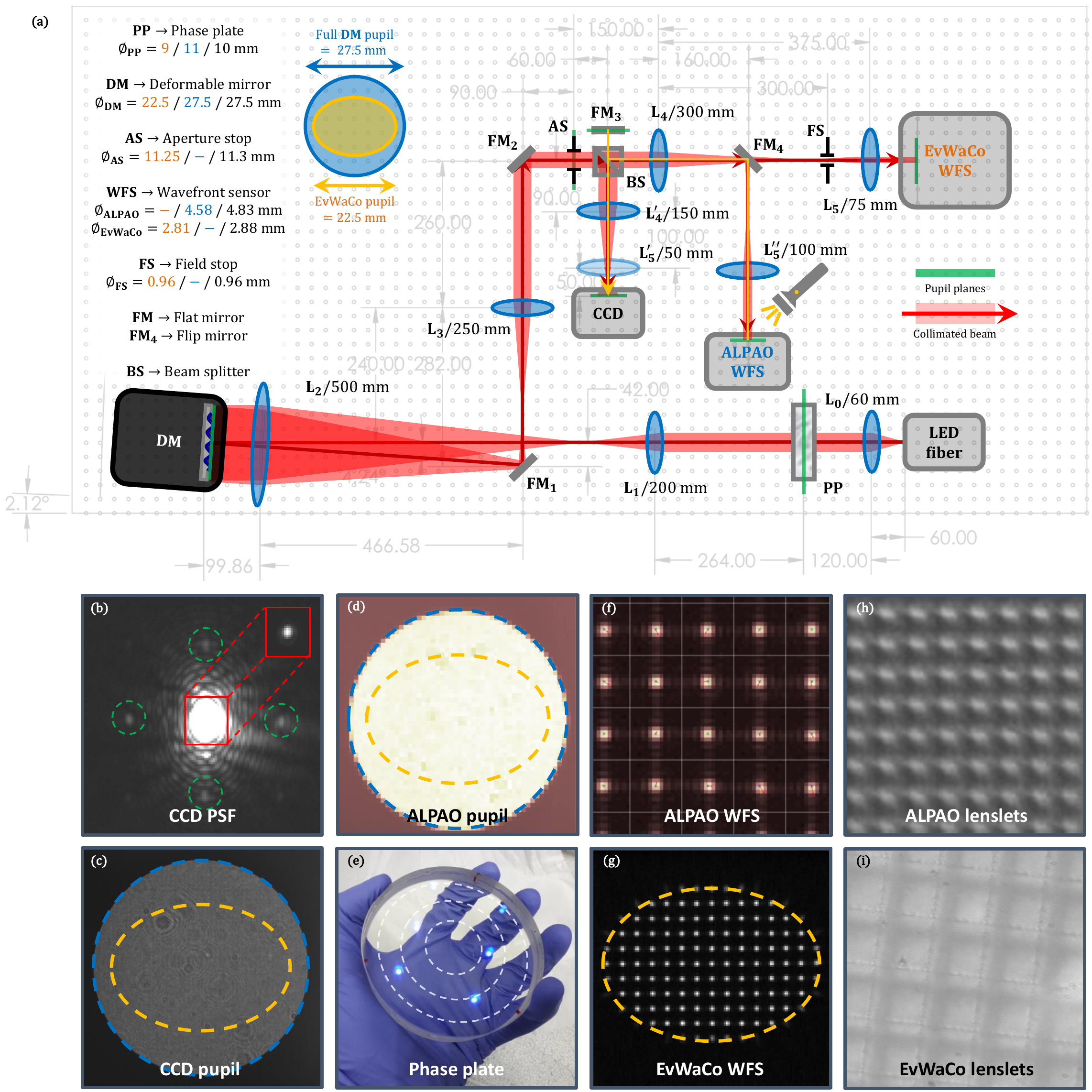}
        \caption{\label{fig:AO_bench} Overview of the adaptive optics bench of EvWaCo. (a) Schematic of the bench. Blue and orange values correspond to the beam in the pupil planes for the full circular pupil of the DM and for the designed elliptical pupil of EvWaCo respectively. The black values give the designed values in the bench. (b) PSF on the CCD after flattening the DM via the ALPAO WFS. The green circles emphasise the secondary diffraction pattern due to the DM actuator regular grid. (c) Image of the pupil on the CCD. (d) Pupil image on the ALPAO WFS. (e) Picture of the Lexitek PP. The two tracks are emphasised by the dashed white circles. (f) Crop on the spots of the ALPAO WFS. (g) Raw data of the EvWaCo WFS on the elliptical pupil. (h,i) Picture of the lenslet arrays of the ALPAO~(h) and EvWaCo~(i) WFS.}
\end{figure}

A schematic of the AO bench is given in \subfig{fig:AO_bench}{a}. Except for \Lens{2} that is a \inch{2} lens to collimate the beam at the full DM size, all the lenses are \inch{1} lenses. The light source is a \SI{617}{\nano\meter} LED (bandwidth of $\Delta\lambda=\SI{15}{\nano\meter}$) coupled with a \SI{10}{\micro\meter} single-mode fiber collimated with \Lens{1}. The PP, shown in \subfig{fig:AO_bench}{e}, is inserted in this first collimated beam. The \Lens{_1}/\Lens{_2} pairing matches the beam size to the full pupil of the DM. The DM is slightly tilted so that the beam goes back through \Lens{_2} and is folded with the flat mirror \FM{1} towards \Lens{3} that scales the beam to the size of the aperture stop \textbf{AS}. The beam splitter \textbf{BS} splits the light in two beams. The first beam goes towards the CCD to image the point spread function (PSF) as shown in \subfig{fig:AO_bench}{b}. The second beam goes towards the two WFS. A flip mirror \FM{4} is used to choose which of the WFS is illuminated. If \FM{4} is in the beam, the light is deflected towards a $54\times54$ SH-WFS from ALPAO, \subfig{fig:AO_bench}{f}. Its resolution is high enough to produce a coarse image of the pupil, \subfig{fig:AO_bench}{d}. If \FM{4} is not in the beam, the light directly goes to the EvWaCo WFS, \subfig{fig:AO_bench}{g}. The pairings \Lens{_4}/\Lens{_5} and \Lens{_4}/\Lens{^{''}_{5}} match the beam diameter to their respective WFS. The field stop \textbf{FS} of the EvWaCo WFS is inserted at the focal point between \Lens{_4} and \Lens{_5}.

Let us notice that a lens \Lens{^{'}_{5}} can be inserted in the beam to shape a pupil image on the CCD as shown in \subfig{fig:AO_bench}{c}. Finally, the flat mirror \FM{3} placed above the \textbf{BS} can be used for pupil and lenslet array imaging with back-propagated light, as schemed by the the orange flash light in \subfig{fig:AO_bench}{a} and shown in \subfigs{fig:AO_bench}{h,i}. This is used for the EvWaCo WFS alignment.

In \subfig{fig:AO_bench}{a}, the black values are the designed or theoretical pupil diameters: the PP is designed for a beam of \SI{10}{\milli\meter} ; the DM has a full aperture of \SI{27.5}{\milli\meter} ; the ALPAO WFS has an aperture of \SI{4.83}{\milli\meter} ; the EvWaCo WFS is designed to have an aperture of \SI{2.88}{\milli\meter} ; a \SI{11.3}{\milli\meter} elliptical aperture stop was available in the laboratory ; the field stop of EvWaCo will have a size of \SI{1}{\milli\meter} which implies a custom-made lens \Lens{3}. These pupil sizes were matched at best by pairing the lenses available in the laboratory. Assuming that the theoretical focal lengths of the chosen lenses are correct, the actual pupil sizes in the bench are given by the orange value. Their reference, by design, is the EvWaCo pupil on DM that is limited by the extreme most actuators: \SI{22.5}{\milli\meter}. Let us notice that doing so, it appears that the pupil size on the PP is smaller than the designed value, which will lead to a slightly better atmospheric turbulence than initially planned. For information, the values of the pupil size when the aperture is the full DM itself are given in blue. This only matters for the characterisation of the DM and the PP with the ALPAO WFS. To do so, the aperture stop \textbf{AS} is removed.

\section{PHASE PLATE DESIGN}
\label{sec:PP}

This sections focuses on the phase plate (PP) of EvWaCo: its design and its characterisation. The PP was manufactured by Lexitek\cite{Ebstein:02_Lexitek} from a design provided by NARIT.

\subsection{Kolmogorov Screen Generation}

To assess the performances of an AO loop, it is critical to emulate the atmospheric turbulence in the bench. Different techniques can be used\cite{Jolissaint:06} such as among the most commonly used: the use of the DM itself, the use of a spatial light modulator (SLM), or the use of a rotating PP\cite{Mantravadi:04}. Using the DM is not optimal as by definition, the DM can correct all the injected turbulence. Spatial frequencies higher than the actuator pitch cannot be emulated and it is thus impossible to assess the performance of the loop to correct for the aliasing of the WFS that wraps the high spatial frequencies. Using a SLM can solve this problem, nonetheless, they work for monochromatic and polarised light. This is problematic as we want to use the fact that EvWaCo is a coronagraph that can work with broadband sources. In addition, EvWaCo has a polarisation dependent behaviour so we do not want to limit the turbulence simulation to a single polarisation state. Finally, SLM are quite slow and thus, the real time performances of the loop cannot be measured. As a consequence, we chose to use a rotating PP to emulate the turbulence in the bench.

Turbulence is a statistical phenomenon characterised by its spatial structure function\cite{Roddier:81}
\begin{equation}
	\SF{\pha}\Paren{\Vx,\Vxp} = \AvgT{\Paren{\pha\Paren{\Vx,t} - \pha\Paren{\Vxp,t}}^{2}}
	\,,
\end{equation}
where $\pha\Paren{\Vx,t}$ is the wavefront phase at the 2D location $\Vx$ at time $t$ and $\AvgT{.}$ stands for the temporal average value. In the following, we work in the framework of a stationary Kolmogorov statistic
\begin{equation}
	\SF{\pha}\Paren{\Vx,\Vxp} = 
	\SF{\pha}\Paren{\Norm{\V{\rho} = \Vx - \Vxp}} = 
	6.88\Paren{\rho/\rFried}^{5/3} 
	\,,
\end{equation}
where $\V{\rho}$ is the radial 2D position, $\rho = \Norm{\V{\rho}}$ its norm and $\rFried$ the Fried's parameter of the turbulence strength\cite{Roddier:81}.

\begin{figure}[ht!]
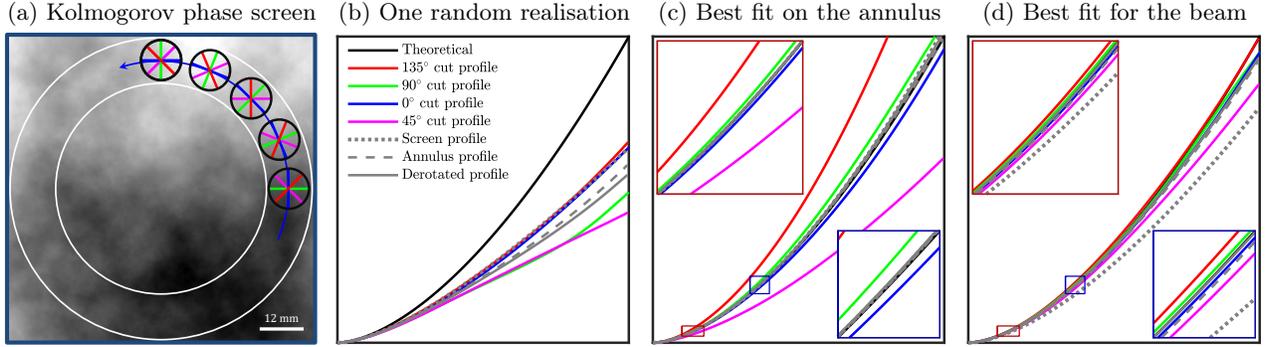
 
        \centering
        
        \newcommand{\LineRatio}{0.95}
        
        \newcommand{\PathFig}{figures_PP_design_}
        \newcommand{\Suffix}{_outer}
        \newcommand{\FigOne}{\PathFig Phase_plate.pdf}
        \newcommand{\FigTwo}{\PathFig profiles_2021_03_23_Lexitek\Suffix .pdf}
        \newcommand{\FigThree}{\PathFig profiles_2021_11_16_NARIT\Suffix .pdf}
        \newcommand{\FigFour}{\PathFig profiles_2021_11_18_NARIT_derot\Suffix .pdf}
        
        \newcommand{\subfigColor}{black}        
        
        \sbox1{\includegraphics{\FigOne}}               
        \sbox2{\includegraphics{\FigTwo}}               
        \sbox3{\includegraphics{\FigThree}}     
        \sbox4{\includegraphics{\FigFour}}     
        \newcommand{\ColumnWidth}[1]
                {\dimexpr \LineRatio \linewidth * \AspectRatio{#1} / (\AspectRatio{1} + \AspectRatio{2} + \AspectRatio{3} + \AspectRatio{4}) \relax
                }
        \newcommand{\ColumnGap}{\hspace {\dimexpr \linewidth /5 - \LineRatio\linewidth /5 }}

        \begin{tabular}{
                @{\ColumnGap}
                C{\ColumnWidth{1}}
                @{\ColumnGap}
                C{\ColumnWidth{2}}
                @{\ColumnGap}
                C{\ColumnWidth{3}}
                @{\ColumnGap}
                C{\ColumnWidth{4}}
                @{\ColumnGap}
                }
                \small (a) Kolmogorov phase screen &
                \small (b) One random realisation &
                \small (c) Best fit on the annulus &
                \small (d) Best fit for the beam
                \\
                \subfigimg[width=\linewidth,pos=ul,font=\fontfig{\subfigColor}]{}{0.0}{\FigOne} &
                \subfigimg[width=\linewidth,pos=ul,font=\fontfig{\subfigColor}]{}{0.0}{\FigTwo} &
                \subfigimg[width=\linewidth,pos=ul,font=\fontfig{\subfigColor}]{}{0.0}{\FigThree} &
                \subfigimg[width=\linewidth,pos=ul,font=\fontfig{\subfigColor}]{}{0.0}{\FigFour}
        \end{tabular}
        
        \caption{\label{fig:PP_design} Design strategy for the phase plate. (a) Example of a randomly generated Kolmogorov phase screen by Lexitek. The white circles emphasise the edges of the track that is rotating in the beam (black circles). The different coloured lines emphasise orientations that are fixed from the beam point of view. (b-d) Radial average (gray) and cut profiles (colors according to the orientation) of the structure functions \textbf{in the beam frame} for different generation strategies. For information, the radial average profiles of the structure functions on the full screen and on the annulus \textbf{in the phase plate frame} are given by the dotted-gray and the dashed-gray profiles respectively. The $x$-axis spans from 0 to the diameter of \SI{12}{\milli\meter} in the phase plate frame chosen for the design. The $y$-axis spans from 0 to the maximal value of theoretical structure function value. (b) Single random generation. (c) Best screen among 1500 that gives the best radial average profile on the annulus frame \textbf{in the PP frame}. (d) Best screen among 1500 that gives the best cut profiles \textbf{in the beam frame}.}
\end{figure}

Different techniques can be used to generate random phase screens. The most common one consists in imposing the turbulence power spectrum density in the frequency domain and adding random phase shifts\cite{McGlamery:76_Turb_PSD, McAulay:00}. This is the technique used by Lexitek\cite{Ebstein:02_Lexitek} for their proposed design screen of  $4096\times4096$ pixels, \subfig{fig:PP_design}{a}. For the design of a single phase screen to be printed on a PP, this technique suffers from several drawbacks. First, the power of the low frequencies is underestimated. Second, nothing guarantees that a single realisation will follow the \textbf{average} Kolmogorov statistics. Third, we actually needs that the beam sees \textbf{in average} and \textbf{in its frame} a 2D Kolmogorov statistics as the PP rotates. These problems are shown in \subfig{fig:PP_design}{b}: even in the \textbf{PP frame} (dotted and dashed curves), the proposed design by Lexitek does not follow the expected structure function. It is strongly underestimated, which is equivalent to a smaller turbulence strength. This would lead to optimistic performances from the bench and its AO loop.

To solve these problems, we propose a first solution to get a more realistic phase screen. To solve the lack of energy in the low spatial frequencies, we inject them in the Kolmogorov spectrum via the sub-harmonics method proposed by Lane\cite{Lane:92_Subharmonics}. $25$ sub-harmonics are added and, as suggested by Lane\cite{Lane:92_Subharmonics}, the screens are simulated on a twice bigger grid whose only the central~$4096 \times 4096$ pixels are kept to avoid aliasing. 1500 screens are generated. Among these screens, the one matching at best the radial structure function on the annulus track, delimited by the white circles in \subfig{fig:PP_design}{a}, is selected. The matching criterion is the sum of the absolute value of the difference between the profiles. We did not choose the sum of the squares to enforce a good matching for the small radii. Indeed, they carry the turbulence high frequencies that are critical to emulate nicely in the bench in order to assess the performances of the AO loop. As shown in \subfig{fig:PP_design}{c}, this technique solves the two first problems \textbf{in the PP frame}: the dashed gray curve nicely overlaps with the theoretical profile. The structure function from the beam point of view is computed by extracting and derotating 180 circular patches along the track as emphasised by the black circles in \subfig{fig:PP_design}{a}. It even appears that \textbf{in the beam frame}, the radial average also follows the expected statistic (gray curve). Nonetheless, the third problem is not solved: the cuts along different directions of the 2D structure functions diverge significantly from the theoretical profile.

As a consequence, we refine further the selection strategy: for each of the 1500 phase screens, the derotated 2D structure functions are computed in order to have a selection criterion \textbf{in the PP frame}. The phase screen whose four structure function cut profiles match at best the theory is selected. As shown in \subfig{fig:PP_design}{d}, at the cost of a slight discrepancy with the radial average profile, this technique allows to select a screen whose 2D structure function is very close to the theory along all the different cut profiles.

The zooms in \subfigs{fig:PP_design}{b,c} show that the selection criterion based on the absolute value works as expected: for small spatial distances, the derotated radial average profiles of the structure function are almost undistinguishable from the theoretical black curve. The Fried's parameters at small scales is thus perfectly controlled to the design value. At the largest scale of the structure functions, a small discrepancy with the theory is noticeable. In this example, this corresponds to an error on the scaling of $\rFried$ of only \SI{2.6}{\percent}.

We designed two tracks with this technique to match two typical nights at the Thai National Observatory (TNO). These two tracks were printed by Lexitek on the same PP on two concentric annuli, see \subfig{fig:AO_bench}{e}. The outer track emulates a $\rFried = \SI{7.4}{\centi\meter}$ that corresponds to an average night at TNO (seeing of $\seeing=\SI{1.4}{''}$). The inner track emulates a $\rFried = \SI{10}{\centi\meter}$ that corresponds to the best quartile at TNO (seeing of $\seeing=\SI{1.0}{''}$). These values are based on preliminary measurements performed in collaboration with Durham university\footnote{By convention, $\rFried$ values are given at $\lambda = \SI{500}{\nano\meter}$. The seeing is given by\cite{Dierickx:92} $\seeing = 0.98 \lambda/\rFried$.}. The structure functions of these designed tracks are given in \subfig{fig:PP_charac}{a}. For the PP design, the beam size is set to \SI{10}{\milli\meter} and the criterion to select the phase screen is based on \SI{12}{\milli\meter} circular patches to allow for margins for the alignment.

\subsection{Characterisation of the Lexitek Phase Plate}

The AO bench is used to characterise the PP manufactured by Lexitek with the $54\times54$ ALPAO SH-WFS, as described in \refsec{sec:AO_bench}. It is critical to know the spatial scaling between the different pupil planes as using the ratio of the focal length of the different lenses is not precise enough. The scaling between the PP plane and the DM is calibrated by imaging the pupil plane on the CCD: first, we know that the DM full pupil is \SI{27.5}{\milli\meter} ; then, we insert a standard in the beam at the PP location with known dimensions\footnote{We used the elliptical aperture stop whose dimensions are designed to be $11.3\times\SI{8}{\milli\meter}$}. The scaling between the DM and the ALPAO WFS is calibrated via the measurements of the DM influence functions (see \refsec{sec:DM}) as the actuator pitch is known to be \SI{1.5}{\milli\meter} on the DM membrane.

\begin{figure}[ht!]
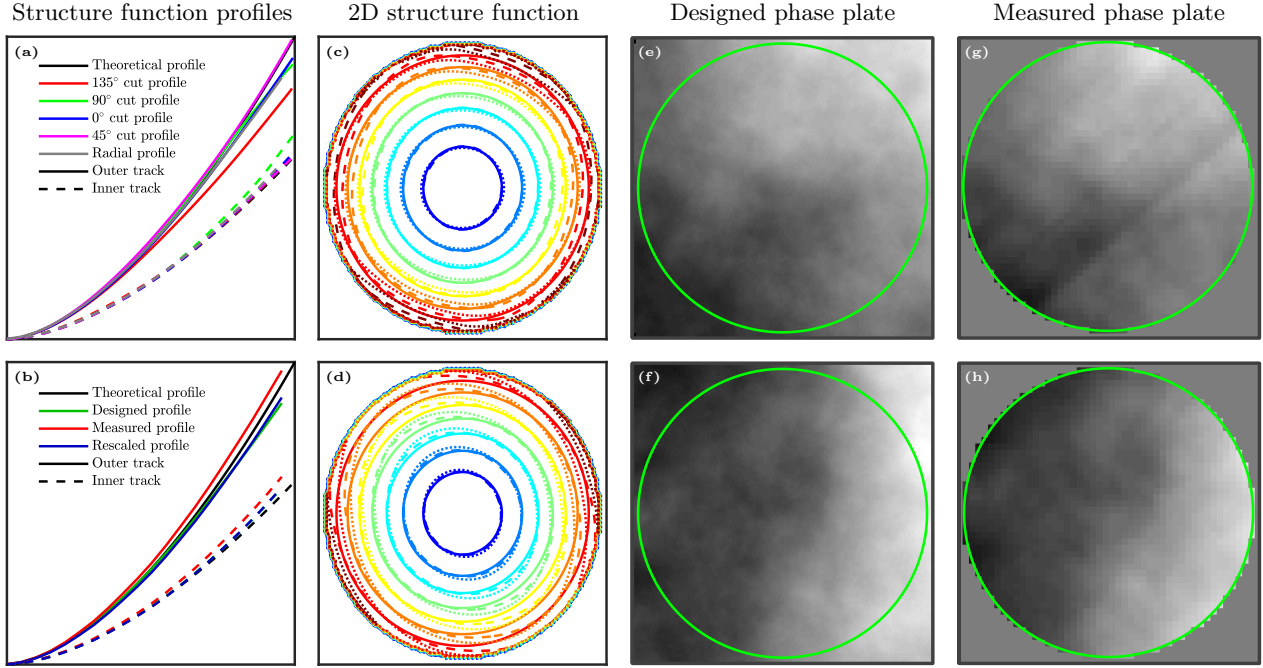
 
        \centering
        
        \newcommand{\LineRatio}{0.95}
        
        \newcommand{\PathFig}{figures_PP_charac_}
        \newcommand{\PrefixOne}{Inner_track_}
        \newcommand{\FigOne}{\PathFig PP_design.pdf}
        \newcommand{\FigTwo}{\PathFig\PrefixOne contour.pdf}
        \newcommand{\FigThree}{\PathFig\PrefixOne Design.pdf}
        \newcommand{\FigFour}{\PathFig\PrefixOne ALPAO.pdf}
        \newcommand{\PrefixTwo}{Outer_track_}
        \newcommand{\FigOneTwo}{\PathFig PP_charac.pdf}
        \newcommand{\FigTwoTwo}{\PathFig\PrefixTwo contour.pdf}
        \newcommand{\FigThreeTwo}{\PathFig\PrefixTwo Design.pdf}
        \newcommand{\FigFourTwo}{\PathFig\PrefixTwo ALPAO.pdf}
        
        \newcommand{\subfigColor}{white}        
        
        \sbox1{\includegraphics{\FigOne}}               
        \sbox2{\includegraphics{\FigTwo}}               
        \sbox3{\includegraphics{\FigThree}}     
        \sbox4{\includegraphics{\FigFour}}     
        \newcommand{\ColumnWidth}[1]
                {\dimexpr \LineRatio \linewidth * \AspectRatio{#1} / (\AspectRatio{1} + \AspectRatio{2} + \AspectRatio{3} + \AspectRatio{4}) \relax
                }
        \newcommand{\ColumnGap}{\hspace {\dimexpr \linewidth /5 - \LineRatio\linewidth /5 }}

        \begin{tabular}{
                @{\ColumnGap}
                C{\ColumnWidth{1}}
                @{\ColumnGap}
                C{\ColumnWidth{2}}
                @{\ColumnGap}
                C{\ColumnWidth{3}}
                @{\ColumnGap}
                C{\ColumnWidth{4}}
                @{\ColumnGap}
                }
                \footnotesize Structure function profiles &
                \footnotesize 2D structure function &
                \footnotesize Designed phase plate &
                \footnotesize Measured phase plate
                \\
                \subfigimg[width=\linewidth,pos=ul,font=\fontfig{black}]{$\;$(a)}{0.0}{\FigOne} &
                \subfigimg[width=\linewidth,pos=ul,font=\fontfig{black}]{$\;$(c)}{0.0}{\FigTwo} &
                \subfigimg[width=\linewidth,pos=ul,font=\fontfig{\subfigColor}]{$\;$(e)}{0.0}{\FigThree} &
                \subfigimg[width=\linewidth,pos=ul,font=\fontfig{\subfigColor}]{$\;$(g)}{0.0}{\FigFour}
                \\
                \subfigimg[width=\linewidth,pos=ul,font=\fontfig{black}]{$\;$(b)}{0.0}{\FigOneTwo} &
                \subfigimg[width=\linewidth,pos=ul,font=\fontfig{black}]{$\;$(d)}{0.0}{\FigTwoTwo} &
                \subfigimg[width=\linewidth,pos=ul,font=\fontfig{\subfigColor}]{$\;$(f)}{0.0}{\FigThreeTwo} &
                \subfigimg[width=\linewidth,pos=ul,font=\fontfig{\subfigColor}]{$\;$(h)}{0.0}{\FigFourTwo}
        \end{tabular}
        
        \caption{\label{fig:PP_charac} Characterisation of the phase plate with the AO bench. (a) Radial (gray) and cut profiles (color) of the design phase plate for the inner (dashed) and outer (plain) tracks. (b) Comparison between the designed (green) and the measured (red) radial profiles. The measurement is rescaled to match at best the design (blue). The $x$-axis spans from 0 to the full diameter of the DM of \SI{27.5}{\milli\meter}. The $y$-axis spans from 0 to the maximal value of theoretical structure function value for the outer track. (c,d) Contour plots of the 2D structure function of the theoretical (plain), designed (dashed) and measured (dotted) structure functions for the inner (c) and outer (d) tracks. The measured structure function is scaled to match at best the designed structure function. (e-g) Comparison between the designed phase plate (e,f) and the measurements (g,h) for a patch the size of a pupil taken in the inner (e,g) and outer (f,h) tracks. The green circles shows the DM pupil in the calibrated phase plate scale.}
\end{figure}

The results are gathered on \fig{fig:PP_charac}. \subfigfull{fig:PP_charac}{b} shows that the PP follows a radial Kolmogorov statistic as expected (red curves). Nonetheless, for both tracks it is slightly higher than the designed tracks (green curves). The best scaling factor is computed for the two tracks to match at best the profiles (blue curves). The spatial scaling corresponds to a correction on $\rFried$ of \SI{-3.4}{\%} for the inner track and \SI{-5.6}{\%} for the outer track. The fact that these factors are similar (negative and same order of magnitude) suggests that they come from systematics. This can be the calibration of the scaling between the different pupil planes, the calibration of the ALPAO SH-WFS, or the material refractive index in the PP. We estimate that these errors are within our bench precision and that the measurements validate the PP design strategy and manufacturing.

\subfigs{fig:PP_charac}{c,d} show that the 2D structure functions of the manufactured tracks match the features of the proposed designs. The small discrepancies are within the precision of our bench.

For information, \subfigs{fig:PP_charac}{e,g} and \subfigs{fig:PP_charac}{f,h} compare a derotated patch of the designed two tracks with its measured counterpart on the manufactured PP. At the low resolution given by the  $54\times54$ ALPAO SH-WFS, the spatial structures are identical. Nonetheless, on the inner track, a band crosses the patch. It is the edge of a square pattern that consequently crosses four times the beam frame during one PP revolution. Its origin is unknown and comes from the PP manufacturing. As shown in \subfigs{fig:PP_charac}{b,c}, it does not impact the structure function significantly and thus this track should still be usable for the AO loop characterisation.

\section{DEFORMABLE MIRROR CHARACTERISATION}
\label{sec:DM}

This section focuses on the deformable mirror (DM) of EvWaCo. It is the DM192 from ALPAO (192 actuators on a circular grid) on which a diaphragm of \SI{27.5}{\milli\meter} has been mounted.

\subsection{Measurement and Data Reduction Procedures}

As for the PP, see \refsec{sec:PP}, the AO bench is used to characterise the DM manufactured by ALPAO with their $54\times54$ SH-WFS, as described in \refsec{sec:AO_bench}. As a side remark, the algorithm provided by ALPAO to measure the slopes from the raw data of the WFS had to be modified to be robust for data acquired with a high dynamic range.

The objective of this section is to measure the influence functions of the DM, that is to say the membrane shape of the DM for each individual actuator when it is poked. In addition, we also want to characterise the linearity of the influence functions, that is to say the similarity of the profile for different amplitude of the command and the proportionality of the profile with the command.

The usual strategy\footnote{That is also the one implemented in the ALPAO Matlab software.} is to push and pull each actuator one by one with a given amplitude. The influence function is the half-difference of the measurements. This method allows getting rid of the constant background phase of the bench. Nonetheless, it is very sensitive to noise as only two measurements are performed for each actuator. In addition, the linearity of the actuator response is not characterised.

\begin{figure}[ht!] 
        \centering
        \includegraphics[width=\linewidth]{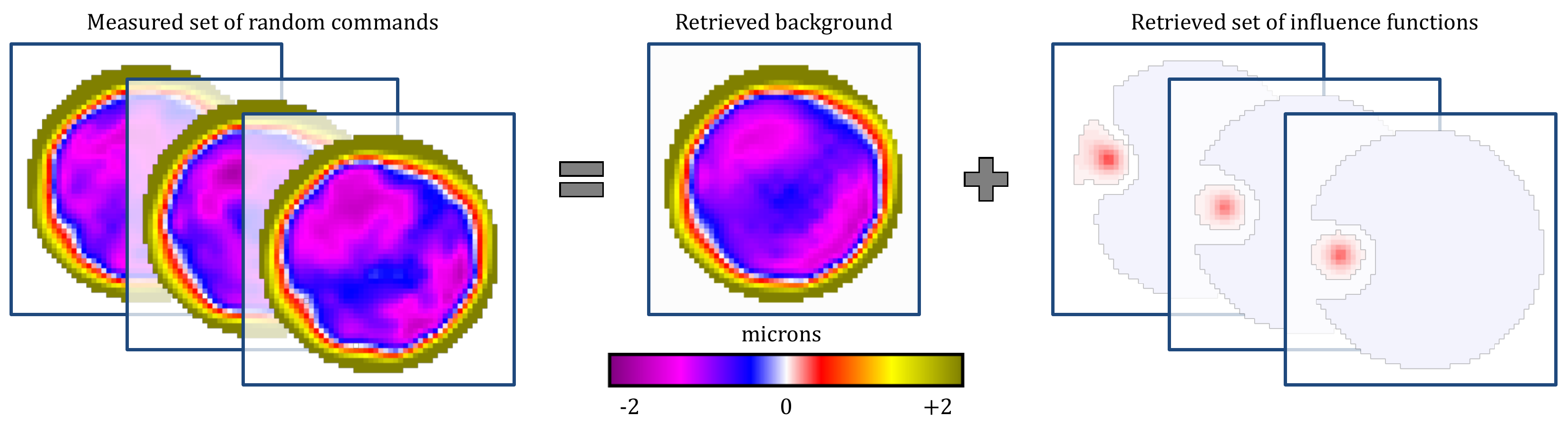}
        \caption{\label{fig:DM_measure} Overview of the measurement strategy to measure the DM influence functions. For a given amplitude, a set of commands where actuators are randomly activated is measured. It is possible to extract from these measurements the overall background of the bench and the influence functions of the DM.}
\end{figure}

In this work, we implement another strategy more robust to noise. For a given command amplitude, a set of binary random (but known) patterns is sent to the DM as schemed in \fig{fig:DM_measure}: for a given pattern, a random list of actuator is poked at the tested command amplitude, the others being left at 0. In the following, we assume that the DM response is linear in terms of influence functions combination\footnote{This hypothesis was not tested.}: when different actuators are poked at the same time, the resulting shape of the DM membrane is the summation of the individual shape obtained when these actuators are poked individually. Thus, the measured phase~$\pham$ at pixel~$\Vpix$ for the $i$-th random pattern can be written
\begin{equation}
	\pham^{i}\Paren{\Vpix} = \phaBG\Paren{\Vpix} + c\times\sum_{a=1}^{\na} \delta^{i}_{a}\IFa\Paren{\Vpix}
	\,,
\end{equation}
where $\na=192$ is the number of actuators, $c$ is the tested command amplitude, $\V{\delta}^{i}\in\Brace{0,1}^{\na}$ is the vector of the $i$-th random binary command, $\IFa$ the influence function of the $a$-th actuator and $\phaBG$ is the background phase of the bench. This equation can be put in a matrix shape:
\begin{equation}
	\Paren{\begin{array}{c}
		\pham^{1}\Paren{\Vpix}
		\\
		\vdots
		\\
		\pham^{\nm}\Paren{\Vpix}
	\end{array}}
	= 
	\Paren{\begin{array}{cccc}
		c\times\delta^{1}_{1} & \cdots & c\times\delta^{1}_{\na} & 1
		\\
		\vdots & \ddots & \vdots & \vdots
		\\
		c\times\delta^{\nm}_{1} & \cdots & c\times\delta^{\nm}_{\na} & 1
	\end{array}}
	\times
	\Paren{\begin{array}{c}
		\pha_{1}\Paren{\Vpix}
		\\
		\vdots
		\\
		\pha_{\na}\Paren{\Vpix}
		\\
		\phaBG\Paren{\Vpix}
	\end{array}}
	\,,
\end{equation}
where $\nm$ is the number of measurements, typically a few times the number of actuators $\na$. Finally, this equation can be inverted to obtain the set of the influence function and background value at the position~$\Vpix$
\begin{equation}
	\Paren{\begin{array}{c}
		\pha_{1}\Paren{\Vpix}
		\\
		\vdots
		\\
		\pha_{\na}\Paren{\Vpix}
		\\
		\phaBG\Paren{\Vpix}
	\end{array}}
	=
	\Paren{\begin{array}{cccc}
		c\times\delta^{1}_{1} & \cdots & c\times\delta^{1}_{\na} & 1
		\\
		\vdots & \ddots & \vdots & \vdots
		\\
		c\times\delta^{\nm}_{1} & \cdots & c\times\delta^{\nm}_{\na} & 1
	\end{array}}\Inv
	\times
	\Paren{\begin{array}{c}
		\pham^{1}\Paren{\Vpix}
		\\
		\vdots
		\\
		\pham^{\nm}\Paren{\Vpix}
	\end{array}}
	\,,
\end{equation}
where $\Paren{.}\Inv$ is the pseudo-inverse of the matrix. This methods is more robust to noise as all the $\nm$ measurements are used in each influence function retrieval. The background phase is also robustly retrieved. This method is schematically described in \fig{fig:DM_measure}. This measurement strategy is applied for the different values of the command $c$ that we want to test.

Now that the 2D maps of the influence functions $\IFa$ have been retrieved for a given command~$c$, the profile of the influence function can be further analysed. By definition, the retrieved influence functions have a 0-piston value on the measured pupil. This means that the background value~$\beta_{a}$ of the actuator~$a$ is slightly negative and this negative value depends on the actuator: the ones on the edge will have a different value than the ones at the center as they depend on the shape of the influence function. In the following we assume that the influence functions~$\IFa$ are axisymmetric and can be described with a normalised radial profile~$\IFrada$
\begin{equation}
	\IFa\Paren{\Vpix} = \alpha_{a}\times\IFrada\paren{\rho_{a}=\Norm{\Vpix-\Vpix_{a}}}+\beta_{a}
	\,,
\end{equation}
where~$\alpha_{a}$ is the actuator profile amplitude, $\Vpix_{a}$ is the 2D position of the $a$-th actuator and~$\beta_{a}$ is the offset insuring a 0-piston value on the measured pupil. In general, a Gaussian profile\cite{Delacroix:15} or a modified Gaussian profile\cite{Jiang:91, Gonzalez-Nunez:17} is fitted. To keep the problem the most general as we can, we describe the profile with splines whose derivative at the most extreme knots is 0. Indeed, we place the first knot at the center of the profile $\rho_{a}=0$ and we place the last knot at 8 actuator pitches. For the first knot the axisymmetry and the smoothness of the influence function impose a 0-derivative. The last knot is far enough from the actuator position to assume that the actuator has no influence on the membrane at this location and that the knot lies in the flat background, also imposing a 0-derivative.

\begin{figure}[ht!]
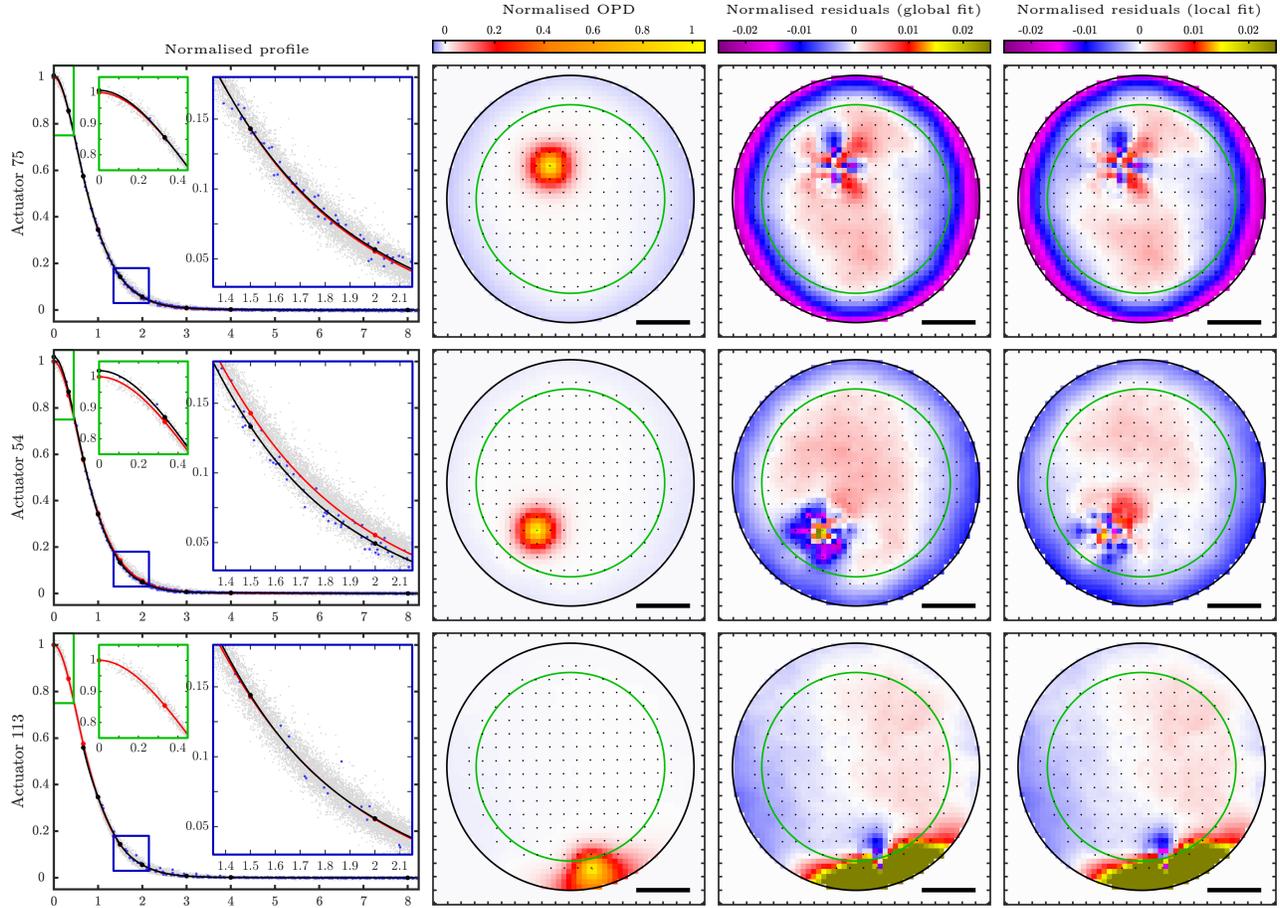
 
        \centering
        
        \newcommand{\LineRatio}{0.95}
        
        \newcommand{\PathFig}{figures_DM_IF_}
        
        \newcommand{\ExtFig}{.png}
        
        \newcommand{\SuffixBar}{_bar}
        \newcommand{\SuffixOne}{_actu_75_C_p040}
        \newcommand{\SuffixTwo}{_actu_54_C_p040}
        \newcommand{\SuffixThree}{_actu_113_C_p040}
        
        \newcommand{\FigOneBar}{\PathFig OPD_rad\SuffixBar \ExtFig}
        \newcommand{\FigTwoBar}{\PathFig OPD\SuffixBar \ExtFig}
        \newcommand{\FigThreeBar}{\PathFig OPD_res_glo\SuffixBar \ExtFig}
        \newcommand{\FigFourBar}{\PathFig OPD_res_loc\SuffixBar \ExtFig}
        
        \newcommand{\FigOneOne}{\PathFig OPD_rad\SuffixOne \ExtFig}
        \newcommand{\FigTwoOne}{\PathFig OPD\SuffixOne \ExtFig}
        \newcommand{\FigThreeOne}{\PathFig OPD_res_glo\SuffixOne \ExtFig}
        \newcommand{\FigFourOne}{\PathFig OPD_res_loc\SuffixOne \ExtFig}

        \newcommand{\FigOneTwo}{\PathFig OPD_rad\SuffixTwo \ExtFig}
        \newcommand{\FigTwoTwo}{\PathFig OPD\SuffixTwo \ExtFig}
        \newcommand{\FigThreeTwo}{\PathFig OPD_res_glo\SuffixTwo \ExtFig}
        \newcommand{\FigFourTwo}{\PathFig OPD_res_loc\SuffixTwo \ExtFig}

        \newcommand{\FigOneThree}{\PathFig OPD_rad\SuffixThree \ExtFig}
        \newcommand{\FigTwoThree}{\PathFig OPD\SuffixThree \ExtFig}
        \newcommand{\FigThreeThree}{\PathFig OPD_res_glo\SuffixThree \ExtFig}
        \newcommand{\FigFourThree}{\PathFig OPD_res_loc\SuffixThree \ExtFig}

        \newcommand{\subfigColor}{black}        
        
        \sbox1{\includegraphics{\FigOneOne}}               
        \sbox2{\includegraphics{\FigTwoOne}}               
        \sbox3{\includegraphics{\FigThreeOne}}     
        \sbox4{\includegraphics{\FigFourOne}}     
        \newcommand{\ColumnWidth}[1]
                {\dimexpr \LineRatio \linewidth * \AspectRatio{#1} / (\AspectRatio{1} + \AspectRatio{2} + \AspectRatio{3} + \AspectRatio{4}) \relax
                }
        \newcommand{\ColumnGap}{\hspace {\dimexpr \linewidth /5 - \LineRatio\linewidth /5 }}

        \begin{tabular}{
                @{\ColumnGap}
                C{\ColumnWidth{1}}
                @{\ColumnGap}
                C{\ColumnWidth{2}}
                @{\ColumnGap}
                C{\ColumnWidth{3}}
                @{\ColumnGap}
                C{\ColumnWidth{4}}
                @{\ColumnGap}
                }
                &
                \tiny Normalised OPD &
                \tiny Normalised residuals (global fit) &
                \tiny Normalised residuals (local fit)
                \\
                \tiny $\quad\quad$ Normalised profile &
                \subfigimg[width=\linewidth,pos=ul,font=\fontfig{\subfigColor}]{}{0.0}{\FigTwoBar} &
                \subfigimg[width=\linewidth,pos=ul,font=\fontfig{\subfigColor}]{}{0.0}{\FigThreeBar} &
                \subfigimg[width=\linewidth,pos=ul,font=\fontfig{\subfigColor}]{}{0.0}{\FigFourBar}
                \\
                \subfigimg[width=\linewidth,pos=ul,font=\fontfig{\subfigColor}]{}{0.0}{\FigOneOne} &
                \subfigimg[width=\linewidth,pos=ul,font=\fontfig{\subfigColor}]{}{0.0}{\FigTwoOne} &
                \subfigimg[width=\linewidth,pos=ul,font=\fontfig{\subfigColor}]{}{0.0}{\FigThreeOne} &
                \subfigimg[width=\linewidth,pos=ul,font=\fontfig{\subfigColor}]{}{0.0}{\FigFourOne}
                \\
                \subfigimg[width=\linewidth,pos=ul,font=\fontfig{\subfigColor}]{}{0.0}{\FigOneTwo} &
                \subfigimg[width=\linewidth,pos=ul,font=\fontfig{\subfigColor}]{}{0.0}{\FigTwoTwo} &
                \subfigimg[width=\linewidth,pos=ul,font=\fontfig{\subfigColor}]{}{0.0}{\FigThreeTwo} &
                \subfigimg[width=\linewidth,pos=ul,font=\fontfig{\subfigColor}]{}{0.0}{\FigFourTwo}
                \\
                \subfigimg[width=\linewidth,pos=ul,font=\fontfig{\subfigColor}]{}{0.0}{\FigOneThree} &
                \subfigimg[width=\linewidth,pos=ul,font=\fontfig{\subfigColor}]{}{0.0}{\FigTwoThree} &
                \subfigimg[width=\linewidth,pos=ul,font=\fontfig{\subfigColor}]{}{0.0}{\FigThreeThree} &
                \subfigimg[width=\linewidth,pos=ul,font=\fontfig{\subfigColor}]{}{0.0}{\FigFourThree}
        \end{tabular}
        
        \caption{\label{fig:DM_IF} Fitting of the influence function radial profiles. Example on three different actuators. Command: $c=+40\%$. All the distances are given in terms of actuator pitch. Scale bar: 4 actuator pitches. \textit{First column --} radial profiles. Gray points: normalised OPD value of all the actuators. Red curves: global average radial profile fit~$\IFradnot$ of the influence functions (the red dots are the knots the spline). Blue points: normalised OPD value of the given actuator. Black curves: radial profile fit of the given actuator influence function~$\IFrada$ (the black dots are the knots of the spline). \textit{Second column --} Normalised OPD map of the given actuator. The black circle is the DM diaphragm. The green circle represents the disk on which the DM performances are guaranteed and on which all the fits are performed. The black dots are the actuator positions. \textit{Third column --} Map of the residuals~$\IFa-\IFradnot$ between the normalised OPD~$\IFa$ of the given actuator with the global fit profile~$\IFradnot$ of the influence function. \textit{Fourth column --} Map of the residuals~$\IFa-\IFrada$ between the normalised OPD~$\IFa$ of the given actuator with the local fit profile~$\IFrada$ of its influence function.}
\end{figure}

There are now two different options. The first option is to assume that all the actuators share the same normalised profile~$\IFradnot$ and only their amplitude~$\alpha_{a}$ and~$\beta_{a}$ are different. By definition,~$\IFradnot$ is normalised between 0 and 1. The second option is to add more freedom by allowing each actuator to have its own radial profile~$\IFrada$. This is done by the Algorithm~\ref{alg:IF_analysis} where the fit of the spline knots is alternated with the fit of the position, amplitude and offset of the actuators. $n_{\Tag{cal}}=15$ loops are performed. To avoid any problem due to outliers on the pupil edge where the DM membrane is fixed with the DM structure, such as seen in \fig{fig:DM_IF} for the actuator 113, only the pixels within a circular mask~$\mask$ centred on the pupil and with a radius of 7 actuator pitches are used for the fit. This is the green circle on \fig{fig:DM_IF}. The amplitude and offset of the actuators are given by the global fit and are fixed for the fit of the individual profiles~$\IFrada$. This means that the local profile can peak higher or lower than 1, as clearly seen in \fig{fig:DM_IF} for the actuator 54. Once the position of the actuators has been finely determined, it is possible to fit the best regular Cartesian grid, given by the black dots on the figure. All the fits are solved with the simplex search method of Lagarias\cite{Lagarias:98_fminsearch}.

\begin{algorithm}
\caption{\label{alg:IF_analysis} Algorithm to retrieve the influence function profiles~$\IFradnot$ and~$\IFrad_{a\in\Bbrack{1,\na}}$ for a given command~$c$.}
\begin{algorithmic}[1]
\small

\For{$a$ from $1$ to $\na$}
	\commentalgo{Initialisation of $\alpha_{a}$, $\beta_{a}$ and $\Vpix_{a}$.}
	
	\State $\Paren{\alpha_{a}, \beta_{a}, \Vpix_{a}, \sigma_{a}} \gets \text{argmin} \sum_{\Vpix \in \mask}\Paren{\alpha_{a}\exp\Paren{-\frac{1}{2\sigma_{a}^{2}}\Norm{\Vpix - \Vpix_{a}}^{2}} + \beta_{a} - \IFa\Paren{\Vpix}}^{2}$
		\commentalgo{Fitting a 2D Gaussian pattern on~$\IFa$.}
\EndFor 

\For{$c$ from $1$ to $n_{\Tag{cal}}$}
	\commentalgo{Several sequential fit of~$\IFradnot$. $\alpha_{a}$, $\beta_{a}$ and $\Vpix_{a}$ are fixed.}
	
	\State $\Tag{\textbf{knot}}_{0} \gets \text{argmin} \sum_{a = 1}^{\na} \sum_{\Vpix \in \mask}\Paren{\alpha_{a}\IFradnot\Paren{\Tag{\textbf{knot}}_{0},\Norm{\Vpix - \Vpix_{a}}} + \beta_{a} - \IFa\Paren{\Vpix}}^{2}$
		\commentalgo{Fitting the knots of the global average profile~$\IFradnot$.}
		
	\State $\Tag{\textbf{knot}}_{0} \gets \Paren{\Tag{\textbf{knot}}_{0} - \Tag{knot}_{0}\Paren{\Tag{last}}}/\Paren{\Tag{knot}_{0}\Paren{\Tag{first}} - \Tag{knot}_{0}\Paren{\Tag{last}}}$
		\commentalgo{Normalisation of the global average profile~$\IFradnot$ between 0 and 1.}
		
	\For{$a$ from $1$ to $\na$}
		\commentalgo{Refining $\alpha_{a}$, $\beta_{a}$ and $\Vpix_{a}$ for each actuator.}
	
		\State $\Paren{\alpha_{a}, \beta_{a}} \gets \text{argmin} \Paren{\alpha_{a}\IFradnot\Paren{\Tag{\textbf{knot}}_{0},\Norm{\Vpix - \Vpix_{a}}} + \beta_{a} - \IFa\Paren{\Vpix}}^{2}$
		\commentalgo{Refining $\alpha_{a}$, $\beta_{a}$. $\Vpix_{a}$ and $\Tag{\textbf{knot}}_{0}$ are fixed.}
		
		\State $\Vpix_{a} \gets \text{argmin} \Paren{\alpha_{a}\IFradnot\Paren{\Tag{\textbf{knot}}_{0},\Norm{\Vpix - \Vpix_{a}}} + \beta_{a} - \IFa\Paren{\Vpix}}^{2}$
		\commentalgo{Refining $\Vpix_{a}$. $\alpha_{a}$, $\beta_{a}$ and $\Tag{\textbf{knot}}_{0}$ are fixed.}
		
	\EndFor 
		
\EndFor

\For{$a$ from $1$ to $\na$}
	\commentalgo{Fitting the individual profiles $\IFrad_{a\in\Bbrack{1,\na}}$. $\alpha_{a}$, $\beta_{a}$ and $\Vpix_{a}$ are fixed.}
	
	\State $\Tag{\textbf{knot}}_{a} \gets \text{argmin} \sum_{\Vpix \in \mask}\Paren{\alpha_{a}\IFrada\Paren{\Tag{\textbf{knot}}_{a},\Norm{\Vpix - \Vpix_{a}}} + \beta_{a} - \IFa\Paren{\Vpix}}^{2}$
		\commentalgo{Fitting the knots of the local profile~$\IFrada$.}
		
\EndFor 

\State \textbf{return} $\Tag{\textbf{knot}}_{0}$ of~$\IFradnot$ and $\Tag{\textbf{knot}}_{a\in\Bbrack{1,\na}}$ of~$\IFrad_{a\in\Bbrack{1,\na}}$
	\commentalgo{Returning the fitted profiles.}
\end{algorithmic}
\end{algorithm}

\figfull{fig:DM_IF} shows the result of the fit for different actuators. The gray points are the normalised optical path difference (OPD) of the measured influence functions~$\Paren{\IFa-\beta_{a}}/\alpha_{a}$ of all the actuators~$a\in\Bbrack{1,\na}$ in which the global profile~$\IFradnot$ is fitted (red curve). The blue points and black curve correspond to the fit on the given actuator~$\Paren{\IFa-\beta_{a}}/\alpha_{a}$ and~$\IFrada$. The actuator 75 has a profile that matches closely the global profile. As a consequences, the residual maps look similar. The actuator 54 is an example of an actuator whose profile~$\IFrada$ is slightly different from the global fit~$\IFradnot$. The profile is thinner than the average, and this can be seen comparing the two residual maps. The actuator 113 is an example of an actuator on the edge of the pupil. Outside the fitted area delimited by the green circle, it can be seen that the axisymmetric hypothesis fails because the membrane is fixed to the DM structure. Nonetheless, within the fitted area, the residual maps for the global and local fit show that this assumption is met on this region, with a profile similar to the average profile.

Finally, the actuators 75 and 54 show that there is a measurable coupling with the neighboring actuators which slightly pull back the membrane. Nonetheless, all these actuators show that the hypothesis of an axisymmetric radial profile is correct within less than a percent (blue and red colors in the figure for the local fit residuals). A finer 2D model similar to the one developed by Huang\cite{Huang:08} could be implemented but was beyond our need and the scope of this work.

We recall here that this fitting strategy is applied for each tested different command~$c$. Thus for each command~$c$, we fit the amplitudes, offsets, positions, radial local and global profiles of the actuators. In the following, for a given command~$c$, they will be noted for~$a\in\Bbrack{1,\na}$:~$\alpha_{a}^{c}$, $\beta_{a}^{c}$, $\Vpix_{a}^{c}$, $\IFrada^{c}$ and $\IFradnot^{c}$.

As a side note, our measurements were cross-validated with the influence function kindly provided by ALPAO at the DM delivery. Despite the lower resolution of the $54\times54$ SH-WFS that we are using, we found the same influence function profiles and amplitudes. This shows that our bench and measurement strategy are quantitative.

\subsection{Characterisation of the ALPAO DM192}

In this section, we analyse the influence function profiles fitted in the previous section to assess their linearity according to the command. We do not perform hysteresis analysis\cite{Jackson:20} and we do not study the DM linearity in terms of influence function combination\cite{Plimmer:18} as this hypothesis was assumed to retrieve the influence function profiles~$\IFa$. Such studies could be made from the obtained measurements but are beyond the scope of this work.

\begin{figure}[ht!]
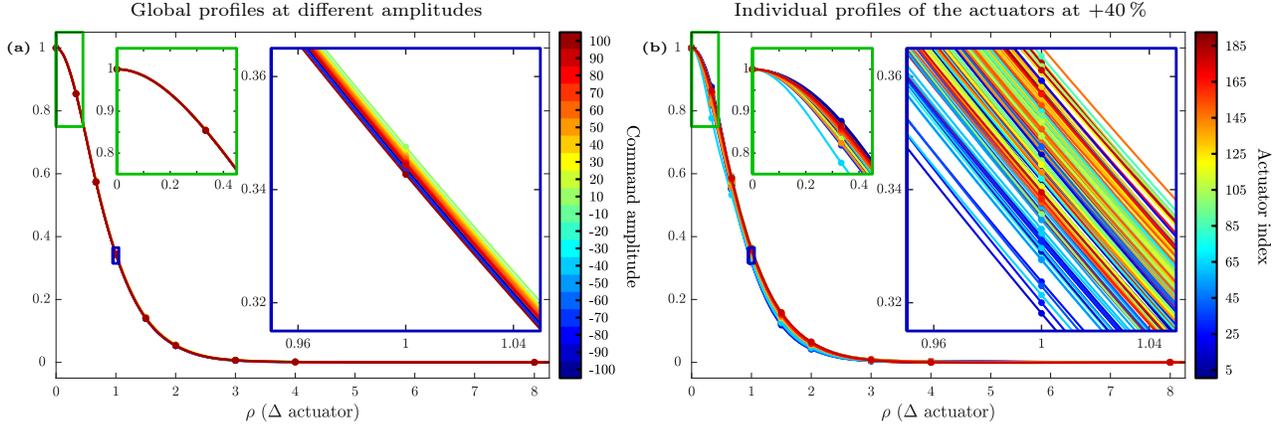
 
        \centering
        
        \newcommand{\LineRatio}{0.95}
        
        \newcommand{\PathFig}{figures_DM_charac_}
        
        \newcommand{\PathSub}{2021_08_30_x8_th50_}
        
        \newcommand{\Suffix}{_C_p040}
        
        \newcommand{\ExtFig}{.pdf}
        
        \newcommand{\FigOne}{\PathFig \PathSub Global_profile\ExtFig}
        \newcommand{\FigTwo}{\PathFig \PathSub Diversity_profile\Suffix \ExtFig}

        \newcommand{\subfigColor}{black}        
        
        \sbox1{\includegraphics{\FigOne}}               
        \sbox2{\includegraphics{\FigTwo}}               
        \newcommand{\ColumnWidth}[1]
                {\dimexpr \LineRatio \linewidth * \AspectRatio{#1} / (\AspectRatio{1} + \AspectRatio{2}) \relax
                }
        \newcommand{\ColumnGap}{\hspace {\dimexpr \linewidth /3 - \LineRatio\linewidth /3 }}

        \begin{tabular}{
                @{\ColumnGap}
                C{\ColumnWidth{1}}
                @{\ColumnGap}
                C{\ColumnWidth{2}}
                @{\ColumnGap}
                }
                \hspace{-0.8cm}\scriptsize Global profiles at different amplitudes &
                \hspace{-0.8cm}\scriptsize Individual profiles of the actuators at $+\SI{40}{\%}$
                \\
                \subfigimg[width=\linewidth,pos=ul,font=\fontfig{\subfigColor}]{\hspace{-11pt}(a)}{0.0}{\FigOne} &
                \subfigimg[width=\linewidth,pos=ul,font=\fontfig{\subfigColor}]{\hspace{-11pt}(b)}{0.0}{\FigTwo}
        \end{tabular}

        \caption{\label{fig:DM_profile} Analysis of the radial profiles of the actuator influence function. (a)~Comparison of the global profiles~$\IFradnot^{c}$ fitted for the different command amplitudes $c\in\Brace{-\SI{100}{\%},+\SI{100}{\%}}$. (b)~Comparison of the local profiles~$\IFrada^{+\SI{40}{\%}}$ fitted for each actuator at the amplitude command $c=+\SI{40}{\%}$.}
\end{figure}

\subfigfull{fig:DM_profile}{a} compares the average profile~$\IFradnot^{c}$ of the influence functions for different commands~$c$. It appears that the coupling (amplitude at 1 inter-actuator distance) is $\sim\SI{34.5}{\%}$. This value slightly changes with the command amplitude. For strong absolute value, the profile gets thinner than around the 0-command. This phenomenon is nonetheless marginal, being below $\SI{0.5}{\%}$ of the actuator amplitude.

\subfigfull{fig:DM_profile}{b} compares the individual actuator profile~$\IFrada^{+\SI{40}{\%}}$ for the command $c=+\SI{40}{\%}$ after their normalisation by~$\Tag{knot}_{a}^{+\SI{40}{\%}}\Paren{\Tag{first}}$. This graph shows that the coupling value of the actuators range from $\sim\SI{32}{\%}$ to $\sim\SI{37}{\%}$.

\begin{figure}[ht!]
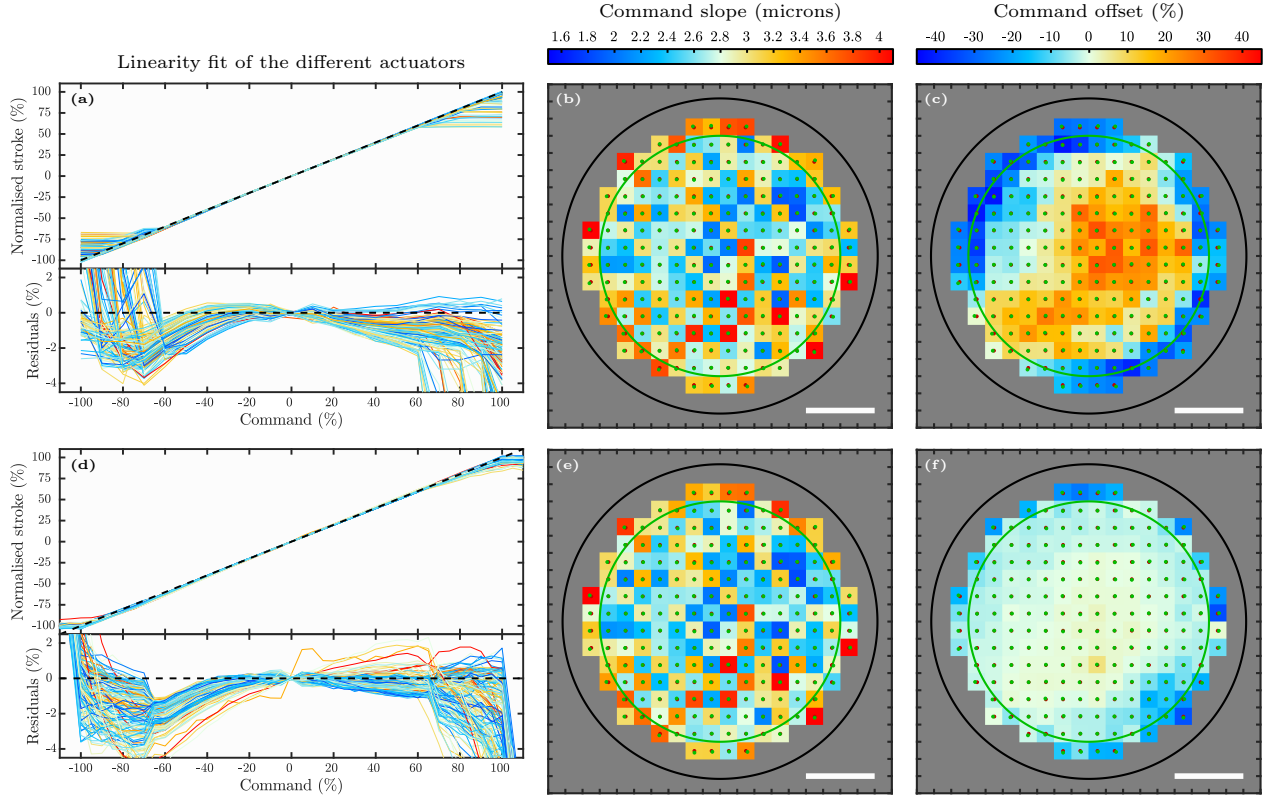
 
        \centering
        
        \newcommand{\LineRatio}{0.95}
        
        \newcommand{\PathFig}{figures_DM_charac_}
        
        \newcommand{\PathSubOne}{2021_08_30_x8_th50_}
        
        \newcommand{\PathSubTwo}{2021_10_21_x5_no_offset_}
        
        \newcommand{\ExtFig}{.pdf}
        
        \newcommand{\FigTwoBar}{\PathFig \PathSubOne Map_slopes_bar\ExtFig}
        \newcommand{\FigThreeBar}{\PathFig \PathSubOne Map_offset_bar\ExtFig}
        
        \newcommand{\FigOneOne}{\PathFig \PathSubOne Linearity_curve\ExtFig}
        \newcommand{\FigTwoOne}{\PathFig \PathSubOne Map_slopes\ExtFig}
        \newcommand{\FigThreeOne}{\PathFig \PathSubOne Map_offset\ExtFig}
        
        \newcommand{\FigOneTwo}{\PathFig \PathSubTwo Linearity_curve\ExtFig}
        \newcommand{\FigTwoTwo}{\PathFig \PathSubTwo Map_slopes\ExtFig}
        \newcommand{\FigThreeTwo}{\PathFig \PathSubTwo Map_offset\ExtFig}

        \newcommand{\subfigColor}{white}        
        
        \sbox1{\includegraphics{\FigOneOne}}               
        \sbox2{\includegraphics{\FigTwoOne}}               
        \sbox3{\includegraphics{\FigThreeOne}}               
        \newcommand{\ColumnWidth}[1]
                {\dimexpr \LineRatio \linewidth * \AspectRatio{#1} / (\AspectRatio{1} + \AspectRatio{2} + \AspectRatio{3}) \relax
                }
        \newcommand{\ColumnGap}{\hspace {\dimexpr \linewidth /4 - \LineRatio\linewidth /4 }}

        \begin{tabular}{
                @{\ColumnGap}
                C{\ColumnWidth{1}}
                @{\ColumnGap}
                C{\ColumnWidth{2}}
                @{\ColumnGap}
                C{\ColumnWidth{3}}
                @{\ColumnGap}
                }
                &
                \scriptsize Command slope (microns) &
                \scriptsize Command offset ($\%$)
                \\
                \scriptsize $\quad\;\;\;$ Linearity fit of the different actuators &
                \subfigimg[width=\linewidth,pos=ul,font=\fontfig{\subfigColor}]{}{0.0}{\FigTwoBar} &
                \subfigimg[width=\linewidth,pos=ul,font=\fontfig{\subfigColor}]{}{0.0}{\FigThreeBar}
                \\
                \subfigimg[width=\linewidth,pos=ul,font=\fontfig{black}]{$\quad\quad\;\;\;$(a)}{0.0}{\FigOneOne} &
                \subfigimg[width=\linewidth,pos=ul,font=\fontfig{\subfigColor}]{$\;$(b)}{0.0}{\FigTwoOne} &
                \subfigimg[width=\linewidth,pos=ul,font=\fontfig{\subfigColor}]{$\;$(c)}{0.0}{\FigThreeOne}
                \\
                \subfigimg[width=\linewidth,pos=ul,font=\fontfig{black}]{$\quad\quad\;\;\;$(d)}{0.0}{\FigOneTwo} &
                \subfigimg[width=\linewidth,pos=ul,font=\fontfig{\subfigColor}]{$\;$(e)}{0.0}{\FigTwoTwo} &
                \subfigimg[width=\linewidth,pos=ul,font=\fontfig{\subfigColor}]{$\;$(f)}{0.0}{\FigThreeTwo}
        \end{tabular}

        \caption{\label{fig:DM_linearity} Analysis of the amplitude linearity of the actuators. (a,d) Response curves of the actuators according to the input command. The actuator response is normalised by their slope~$\alpha_{a}$. The color map corresponds to the color of the command slope of (b,e). (b,e) 2D map of the actuator command slopes~$\V{\slope}$. Scale bar: 4 actuator pitches. Coloured circles: see \fig{fig:DM_IF}. Green points: fitted position~$\Vpix_{a}$ of each actuator. Red points: closest regular grid on the fitted area. (c,f) 2D map of the actuator command offsets~$\V{\offset}$. (a-c) Using the configuration files provided with the DM. (d-f) Using the configuration asked to ALPAO without offset in the command.}    
\end{figure}

It is then possible to study the linearity of the stroke (in microns) with the actuator command. According to the fitting strategy, the measured stroke for a given actuator~$a$ and a given command~$c$ is
\begin{equation}
	\stroke\Paren{a,c} = c\times\alpha^{c}_{a}\times\Tag{knot}^{c}_{a}\Paren{\Tag{first}}
	\,.
\end{equation}
Intuitively, the user expects that the Software Development Kit (SDK) of the DM clips the command between $-\SI{100}{\%}$ and $+\SI{100}{\%}$. But we noticed some unexpected behaviours. As a consequence, we decided to fit a possible offset in this clipping. The slope~$\tilde{\alpha}_{a}$ and the offset~$\offset_{a}$ of each actuator are fitted by solving with the simplex search method of Lagarias\cite{Lagarias:98_fminsearch}
\begin{equation}
	\Paren{\slope_{a}, \offset_{a}} = \text{argmin} \sum_c \Paren{\stroke\Paren{a,c} - \slope_{a}\ClipVal{c-\offset_{a}}{-\SI{100}{\%}}{+\SI{100}{\%}}}^{2}
	\,,
\end{equation}
where $\ClipVal{.}{l_1}{l_2}$ is the function that clips its argument between $l_1$ and $l_2$. The results are gathered in \fig{fig:DM_linearity}.

\subfigsfull{fig:DM_linearity}{a,d} show the normalised strokes~$\stroke\Paren{a,c}/\slope_{a}$ as well as the linearity residuals~$\stroke\Paren{a,c}/\slope_{a}-c$. It can clearly be seen from \subfig{fig:DM_linearity}{a} that unexpected clipping occurs within the expected command $\Brack{-\SI{100}{\%},+\SI{100}{\%}}$ range. This clipping values are consequently seen in the offset map of \subfig{fig:DM_linearity}{c}. We thus concluded that the ALPAO SDK does not clip the user's command~$\ClipVal{c}{-\SI{100}{\%}}{+\SI{100}{\%}}$ but~$\ClipVal{c-\offset_{a}}{-\SI{100}{\%}}{+\SI{100}{\%}}$, that is to say the user's command corrected by an offset~$\V{\offset}$ set by ALPAO but in general not known by the user. This can be highly misleading as this offset range can reach value as high as $+\SI{40}{\%}$ and as low as $-\SI{40}{\%}$ as shown on the figure.

This hypothesis was confirmed by asking ALPAO the list of their offsets for our DM192. Their values matched our measurements: this result shows once again that our bench and measurement strategy are quantitative.

A new configuration file was asked to ALPAO in which the offset map is set to 0 for all the actuators. New measurements were made and are presented in \subfigs{fig:DM_linearity}{d,e,f}. Except for some actuators on the edge beyond the fitting area and an actuator which also seem to have some linearity issues, all the actuators are measured to have a negligible offset in \subfig{fig:DM_linearity}{f}. The not-strictly null values are within the noise of our measurements. The profiles of \subfig{fig:DM_linearity}{d} show that the actuators amplitude is linear within $\Brack{-\SI{4}{\%},+\SI{2}{\%}}$ on the full dynamics of the command $\Brack{-\SI{100}{\%},+\SI{100}{\%}}$.

\subfigsfull{fig:DM_linearity}{b,e} show the maps of the actuator command slope. The median value is \SI{2.8}{\micro\meter}. Some actuators can be beyond a \SI{4}{\micro\meter} stroke. On the upper right, there is nonetheless a cluster of actuators whose stroke is significantly lower than the median value, around \SI{1.8}{\micro\meter}. Side note: as expected, the slope maps~$\V{\slope}$ of \subfigs{fig:DM_linearity}{b,e} are almost identical as this parameter should not be impacted by the value of the offsets.

\begin{figure}[ht!]
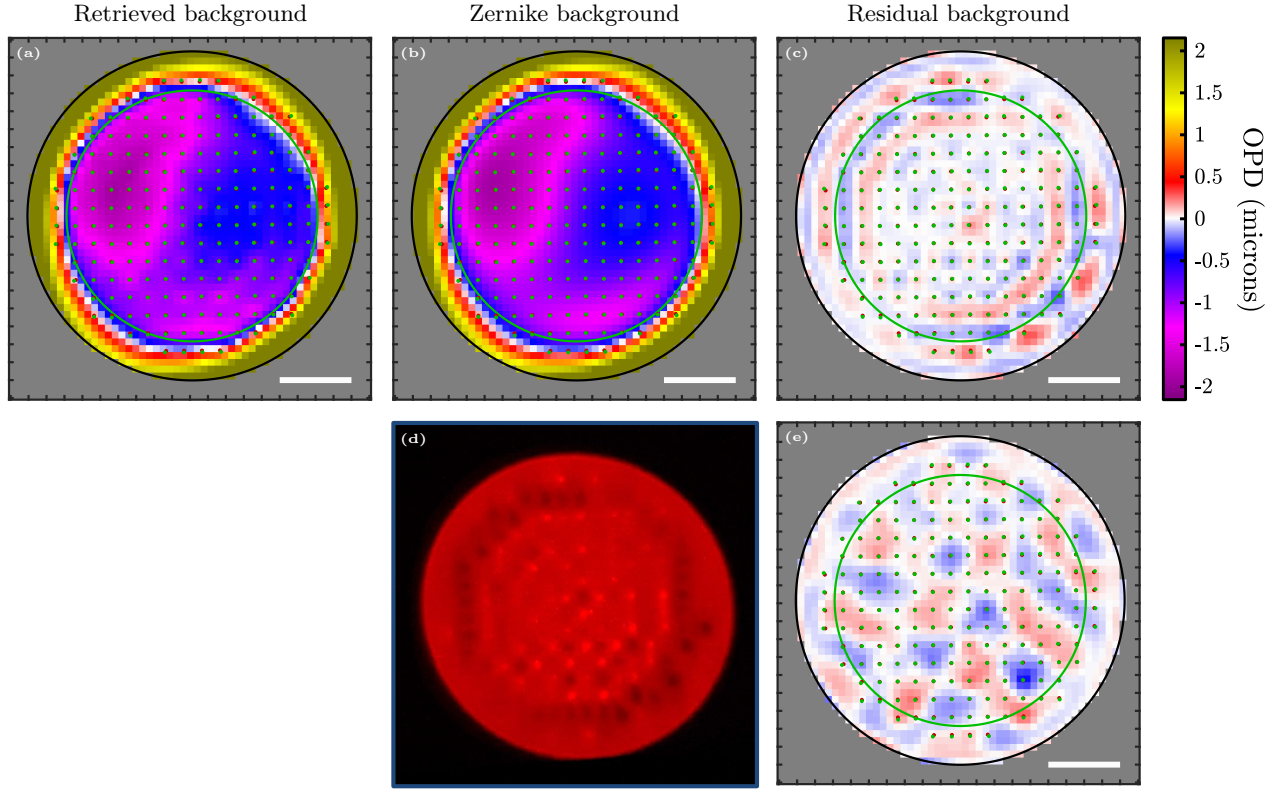
 
        \centering
        
        \newcommand{\LineRatio}{0.95}
        
        \newcommand{\PathFig}{figures_DM_charac_}
        
        \newcommand{\PathSubOne}{2021_08_30_x8_th50_}
        
        \newcommand{\PathSubTwo}{2021_10_21_x5_no_offset_}

        \newcommand{\Suffix}{_C_p040}
        
        \newcommand{\ExtFig}{.pdf}
        
        \newcommand{\FigOneOne}{\PathFig \PathSubOne BG_dat\Suffix \ExtFig}
        \newcommand{\FigTwoOne}{\PathFig \PathSubOne BG_fit\Suffix \ExtFig}
        \newcommand{\FigThreeOne}{\PathFig \PathSubOne BG_res\Suffix \ExtFig}
        \newcommand{\FigFourOne}{\PathFig \PathSubOne Map_BG_bar\ExtFig}
        
        \newcommand{\FigAux}{\PathFig fig_2021_08_30_x8_th50\ExtFig}
        \newcommand{\FigThreeTwo}{\PathFig \PathSubTwo BG_res\Suffix \ExtFig}

        \newcommand{\subfigColor}{white}        
        
        \sbox1{\includegraphics{\FigOneOne}}               
        \sbox2{\includegraphics{\FigTwoOne}}               
        \sbox3{\includegraphics{\FigThreeOne}}               
        \sbox4{\includegraphics{\FigFourOne}}               
        \newcommand{\ColumnWidth}[1]
                {\dimexpr \LineRatio \linewidth * \AspectRatio{#1} / (\AspectRatio{1} + \AspectRatio{2} + \AspectRatio{3} + \AspectRatio{4}) \relax
                }
        \newcommand{\ColumnGap}{\hspace {\dimexpr \linewidth /5 - \LineRatio\linewidth /5 }}

        \begin{tabular}{
                @{\ColumnGap}
                C{\ColumnWidth{1}}
                @{\ColumnGap}
                C{\ColumnWidth{2}}
                @{\ColumnGap}
                C{\ColumnWidth{3}}
                @{\ColumnGap}
                C{\ColumnWidth{4}}
                @{\ColumnGap}
                }
                \footnotesize Retrieved background &
                \footnotesize Zernike background &
                \footnotesize Residual background &
                \\
                \subfigimg[width=\linewidth,pos=ul,font=\fontfig{\subfigColor}]{$\;$(a)}{0.0}{\FigOneOne} &
                \subfigimg[width=\linewidth,pos=ul,font=\fontfig{\subfigColor}]{$\;$(b)}{0.0}{\FigTwoOne} &
                \subfigimg[width=\linewidth,pos=ul,font=\fontfig{\subfigColor}]{$\;$(c)}{0.0}{\FigThreeOne} &
                \subfigimg[width=\linewidth,pos=ul,font=\fontfig{\subfigColor}]{}{0.0}{\FigFourOne}
                \\
                &
                \subfigimg[width=\linewidth,pos=ul,font=\fontfig{\subfigColor}]{$\;$(d)}{0.0}{\FigAux} &
                \subfigimg[width=\linewidth,pos=ul,font=\fontfig{\subfigColor}]{$\;$(e)}{0.0}{\FigThreeTwo}
                &
                \\
        \end{tabular}

        \caption{\label{fig:DM_background} Analysis of the OPD background retrieved from the influence function measurements at the amplitude command $+40\%$. (a) Retrieved background = (b) Zernike fit of the background + (c) Residuals of the background fit using the configuration file provided with the DM. (d) Image obtained on a screen inserted in the beam after the reflection on the DM membrane when the 0-command is applied using the configuration file provided with the DM. (e)  Same as (c) but with a configuration file setting the offsets to 0. Legend of the coloured circles and dots: see \figs{fig:DM_IF}{fig:DM_linearity}.}
\end{figure}

We mentioned in the previous section that the phase background of the bench~$\phaBG$ can be extracted via our fitting strategy. It is shown in \subfig{fig:DM_background}{a}. Some meaningful information can be extracted from this measurement. Indeed, its high-order spatial frequencies are likely coming from the DM itself and a wrong flat command. To split this contribution from the low-order aberrations, we fit 65 Zernike modes in the background image\cite{Fricker:02_Zernike}. The obtained fit is given in \subfig{fig:DM_background}{b} and its residuals in \subfig{fig:DM_background}{c}.

By comparing \subfig{fig:DM_linearity}{c} and \subfig{fig:DM_background}{b}, it appears that the coarse spatial features are identical (blue \vs purple and orange \vs dark blue). This implies that the offsets set by ALPAO on the mirror produce a global low-order aberration that we measure in the background phase. Then, by comparing \subfig{fig:DM_linearity}{c} and \subfig{fig:DM_background}{c}, it appears that the high-order aberrations present in the background phase follow the local gradient in the offset maps. This high-order aberrations of \subfig{fig:DM_background}{c} also nicely match the intensity pattern of \subfig{fig:DM_background}{d} obtained by placing a screen in the beam in a plane slightly out of focus from the pupil position. Doing so, phase aberrations turn into intensity inhomogeneities. This validates the Zernike fit strategy and shows that the bench and the influence function fitting strategy are sensitive to these small artefacts. \subfigfull{fig:DM_background}{e} shows the background residual map obtained with the new configuration where the actuator offsets are all set to 0. The high frequencies at the actuator pitch scale from \subfig{fig:DM_background}{c} are not present any more and the residuals appear to be smooth. This is a reassuring result: it means that there is no strong artefact at a scale lower than the actuator pitch that the DM could not correct. Such artefacts smaller than the actuator pitch have been noticed in our Boston Micromachines Corporation DM which produce strong secondary diffraction patterns, as discussed in \refapp{app:DM_BMC}.

\section{CONCLUSION}

In this work, we introduced the AO system of EvWaCo based on a DM with $\na=192$ actuators and a $15\times11$ sub-apertures SH-WFS. A dedicated bench was designed and aligned in the NARIT cleanroom to develop and study this AO system. One of its objective is to characterise its different critical components.

In this bench, the atmospheric turbulence is emulated by a rotating phase plate. We presented in this work a new method to generate and select a phase screen that ensures that the beam sees a perfect and controlled 2D Kolmogorov structure function. The manufactured device from Lexitek has been characterised with our bench and we confirmed that it produces a Kolomogorov turbulence as it rotates with the correct 2D statistic.

We also used this bench to characterise the ALPAO DM192. We developed a new algorithm to fit its influence function based on a spline axisymmetric profile. We confirmed that the axysimmetric spline model is accurate enough to fit the DM influence function within the order of the percent. The small discrepancies come from some coupling with the neighboring actuators, slightly breaking the axysimmetry assumption. We also showed that there is some dispersion in the actuator profile shape. We noticed that the clipping of the user's command by the SDK is not intuitive and takes into account an offset set by ALPAO but unknown to the user. This problem is easily corrected by creating a new configuration file with the offset all set to 0. We measured some deviations from linearity concerning the stroke in function of the command. Our method also allows extracting the bench phase background. We studied this background to decouple the low-order aberrations coming from the bench and the general shape of the DM from the high-order aberrations close to the actuator pitch scale. We showed that most of the features were coming from the offset map loaded by ALPAO in their configuration file. Starting from a 0-offset map appears to be better.

In the near future, this bench will also be used for the development of the code to control the hardware and run the AO loop. It will allow testing new AO loop algorithms and characterise their performances. They will be compared with a full end-to-end model of the system that has been developed to simulate the bench\cite{Ridsdill:22} as well as the full EvWaCo instrument on a computer\cite{Berdeu:22_AO_loop}.

\appendix    

\section{DEFORMABLE MIRROR CHARACTERISATION}
\label{app:DM_BMC}

In this appendix, we show that the algorithms that were developed for the characterisation of the DM129 from ALPAO are versatile by using them on a Boston Micromachines Corporation (BMC) DM\footnote{\url{https://bostonmicromachines.com}}. It is a $12\times12$ actuator DM on a square grid, without any actuator in the four corners (140 actuators in total). This BMC DM is currently under alignment in the EvWaCo testbed\cite{Alagao:21_EvWaCo_exp_contrast} to correct the static aberrations of the bench and assess the diffraction limited performances of the focal plane mask.

As a side note, the BMC SDK works for commands between 0 and 1. In the following, we work around the average value of 0.5 that is set as a reference, mapping $\Brack{0,1}$ of the SDK onto $\Brack{\SI{-100}{\percent},\SI{100}{\percent}}$.

\begin{figure}[ht!]
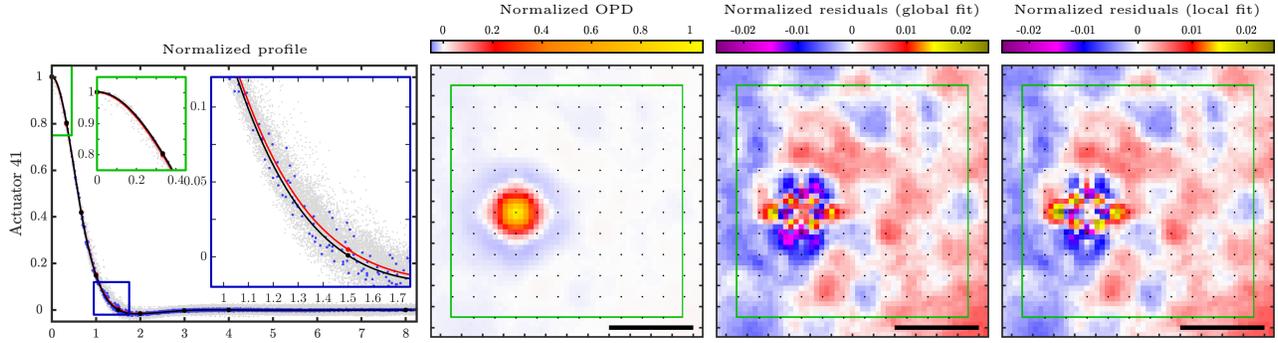
 
        \centering
        
        \newcommand{\LineRatio}{0.95}
        
        \newcommand{\PathFig}{figures_DM_BMC_IF_}
        
        \newcommand{\ExtFig}{.png}
        
        \newcommand{\SuffixBar}{_bar}
        \newcommand{\SuffixOne}{_actu_41_C_p040}

        \newcommand{\FigOneBar}{\PathFig OPD_rad\SuffixBar \ExtFig}
        \newcommand{\FigTwoBar}{\PathFig OPD\SuffixBar \ExtFig}
        \newcommand{\FigThreeBar}{\PathFig OPD_res_glo\SuffixBar \ExtFig}
        \newcommand{\FigFourBar}{\PathFig OPD_res_loc\SuffixBar \ExtFig}
        
        \newcommand{\FigOneOne}{\PathFig OPD_rad\SuffixOne \ExtFig}
        \newcommand{\FigTwoOne}{\PathFig OPD\SuffixOne \ExtFig}
        \newcommand{\FigThreeOne}{\PathFig OPD_res_glo\SuffixOne \ExtFig}
        \newcommand{\FigFourOne}{\PathFig OPD_res_loc\SuffixOne \ExtFig}

        \newcommand{\subfigColor}{black}        
        
        \sbox1{\includegraphics{\FigOneOne}}               
        \sbox2{\includegraphics{\FigTwoOne}}               
        \sbox3{\includegraphics{\FigThreeOne}}     
        \sbox4{\includegraphics{\FigFourOne}}     
        \newcommand{\ColumnWidth}[1]
                {\dimexpr \LineRatio \linewidth * \AspectRatio{#1} / (\AspectRatio{1} + \AspectRatio{2} + \AspectRatio{3} + \AspectRatio{4}) \relax
                }
        \newcommand{\ColumnGap}{\hspace {\dimexpr \linewidth /5 - \LineRatio\linewidth /5 }}

        \begin{tabular}{
                @{\ColumnGap}
                C{\ColumnWidth{1}}
                @{\ColumnGap}
                C{\ColumnWidth{2}}
                @{\ColumnGap}
                C{\ColumnWidth{3}}
                @{\ColumnGap}
                C{\ColumnWidth{4}}
                @{\ColumnGap}
                }
                &
                \tiny Normalized OPD &
                \tiny Normalized residuals (global fit) &
                \tiny Normalized residuals (local fit)
                \\
                \tiny $\quad\quad$ Normalized profile &
                \subfigimg[width=\linewidth,pos=ul,font=\fontfig{\subfigColor}]{}{0.0}{\FigTwoBar} &
                \subfigimg[width=\linewidth,pos=ul,font=\fontfig{\subfigColor}]{}{0.0}{\FigThreeBar} &
                \subfigimg[width=\linewidth,pos=ul,font=\fontfig{\subfigColor}]{}{0.0}{\FigFourBar}
                \\
                \subfigimg[width=\linewidth,pos=ul,font=\fontfig{\subfigColor}]{}{0.0}{\FigOneOne} &
                \subfigimg[width=\linewidth,pos=ul,font=\fontfig{\subfigColor}]{}{0.0}{\FigTwoOne} &
                \subfigimg[width=\linewidth,pos=ul,font=\fontfig{\subfigColor}]{}{0.0}{\FigThreeOne} &
                \subfigimg[width=\linewidth,pos=ul,font=\fontfig{\subfigColor}]{}{0.0}{\FigFourOne}
        \end{tabular}
        
        \caption{\label{fig:DM_BMC_IF} Fitting of the influence function radial profiles of the BMC DM. Example on the actuator 41 for a command amplitude of \SI{40}{\percent}. See caption of \fig{fig:DM_IF}. The green square represents the area on which all the fits are performed.}
\end{figure}

\figfull{fig:DM_BMC_IF} presents the optical path difference (OPD) profile of the DM influence functions. From the radial profile of \subfig{fig:DM_BMC_IF}{a} and the blue ring in the 2D map of \subfig{fig:DM_BMC_IF}{b}, it appears that the influence functions have a non-negligible negative overshoot peaking at a distance of two actuator pitches. It seems from \subfig{fig:DM_BMC_IF}{b}, that the shape of the influence function is not strictly radially symmetric, looking more like a diamond slightly more elongated along the $x$-axis. This is nonetheless negligible as shown by the good radial fit in \subfig{fig:DM_BMC_IF}{a}. The discrepancies between the global residual map of \subfig{fig:DM_BMC_IF}{c} and the local residual map of \subfig{fig:DM_BMC_IF}{d} suggest some dispersion in the influence function individual profiles. The cross-talk with neighboring actuators is also important, maybe explaining the mentioned negative overshoot.

\begin{figure}[ht!]
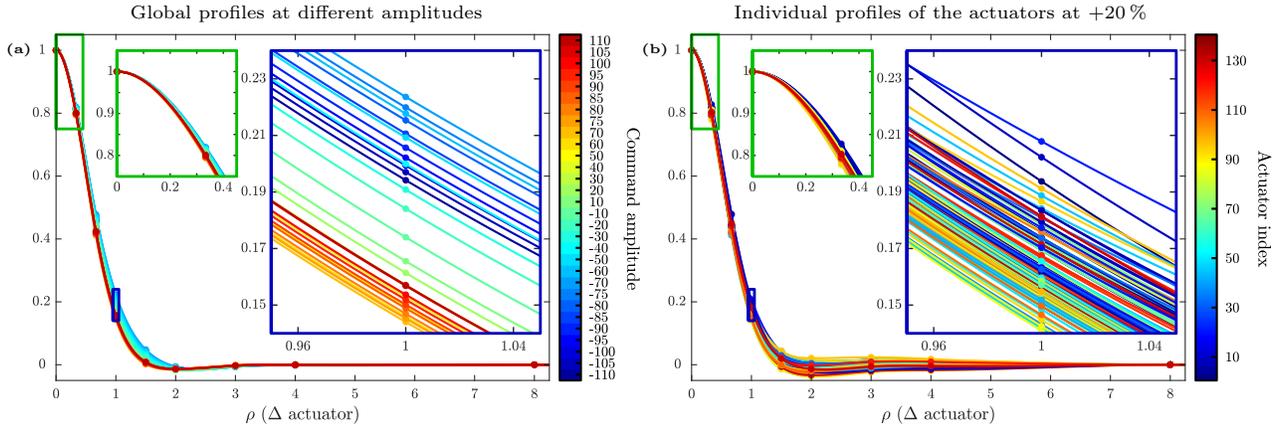
 
        \centering
        
        \newcommand{\LineRatio}{0.95}
        
        \newcommand{\PathFig}{figures_DM_BMC_charac_}
        
        \newcommand{\Suffix}{_C_p020}
        
        \newcommand{\ExtFig}{.pdf}
        
        \newcommand{\FigOne}{\PathFig Global_profile\ExtFig}
        \newcommand{\FigTwo}{\PathFig Diversity_profile\Suffix \ExtFig}

        \newcommand{\subfigColor}{black}        
        
        \sbox1{\includegraphics{\FigOne}}               
        \sbox2{\includegraphics{\FigTwo}}               
        \newcommand{\ColumnWidth}[1]
                {\dimexpr \LineRatio \linewidth * \AspectRatio{#1} / (\AspectRatio{1} + \AspectRatio{2}) \relax
                }
        \newcommand{\ColumnGap}{\hspace {\dimexpr \linewidth /3 - \LineRatio\linewidth /3 }}

        \begin{tabular}{
                @{\ColumnGap}
                C{\ColumnWidth{1}}
                @{\ColumnGap}
                C{\ColumnWidth{2}}
                @{\ColumnGap}
                }
                \hspace{-0.8cm}\scriptsize Global profiles at different amplitudes &
                \hspace{-0.8cm}\scriptsize Individual profiles of the actuators at $+\SI{20}{\percent}$
                \\
                \subfigimg[width=\linewidth,pos=ul,font=\fontfig{\subfigColor}]{\hspace{-11pt}(a)}{0.0}{\FigOne} &
                \subfigimg[width=\linewidth,pos=ul,font=\fontfig{\subfigColor}]{\hspace{-11pt}(b)}{0.0}{\FigTwo}
        \end{tabular}

        \caption{\label{fig:DM_BMC_profile} Analysis of the radial profiles of the actuator influence function of the BMC DM. See caption of \fig{fig:DM_profile}. (a)~$c\in\Brace{-\SI{110}{\%},+\SI{110}{\%}}$. (b)~$c=+\SI{20}{\%}$.}
\end{figure}

The linearity of the influence function profiles with the command amplitude is studied in \fig{fig:DM_BMC_profile}. As seen in \subfig{fig:DM_BMC_profile}{a}, the actuator coupling value of the average influence function profile noticeably depends on the command amplitude, ranging from \SI{\sim 14}{\percent} to \SI{\sim 23}{\percent}. For a given command amplitude of \SI{20}{\percent}, as shown in \subfig{fig:DM_BMC_profile}{b}, the dispersion of the coupling value among the different actuators is also important, ranging from \SI{\sim 13}{\percent} to \SI{\sim 21}{\percent}.

\begin{figure}[ht!]
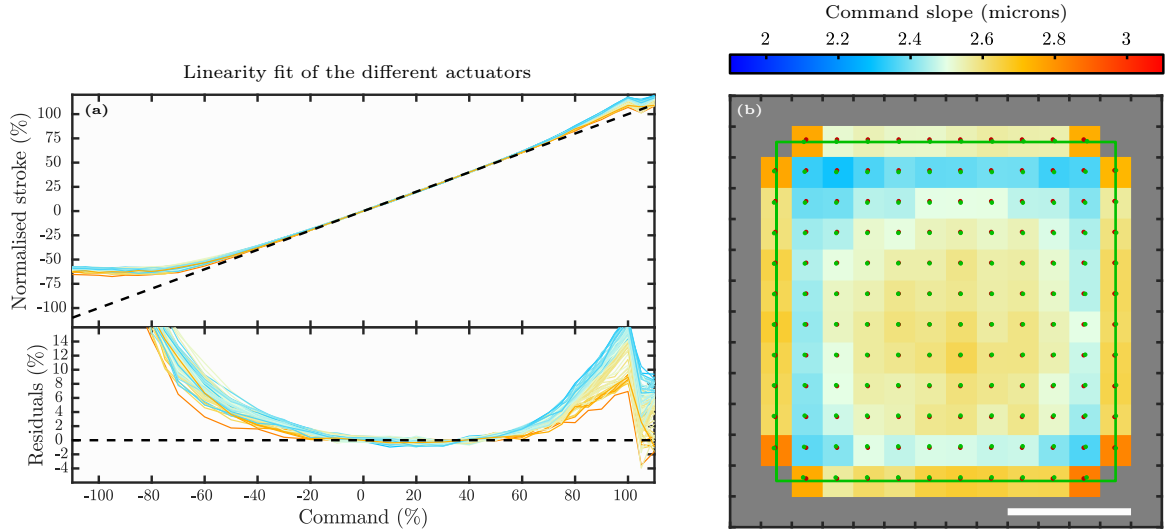
 
        \centering
        
        \newcommand{\LineRatio}{0.85}
        
        \newcommand{\PathFig}{figures_DM_BMC_charac_}
        
        \newcommand{\ExtFig}{.pdf}
        
        \newcommand{\FigTwoBar}{\PathFig Map_slopes_bar\ExtFig}
        
        \newcommand{\FigOneOne}{\PathFig Linearity_curve\ExtFig}
        \newcommand{\FigTwoOne}{\PathFig Map_slopes\ExtFig}

        \newcommand{\subfigColor}{white}        
        
        \sbox1{\includegraphics{\FigOneOne}}               
        \sbox2{\includegraphics{\FigTwoOne}}               

        \newcommand{\ColumnWidth}[1]
                {\dimexpr \LineRatio \linewidth * \AspectRatio{#1} / (\AspectRatio{1} + \AspectRatio{2}) \relax
                }
        \newcommand{\ColumnGap}{\hspace {\dimexpr \linewidth /3 - \LineRatio\linewidth /3 }}

        \begin{tabular}{
                @{\ColumnGap}
                C{\ColumnWidth{1}}
                @{\ColumnGap}
                C{\ColumnWidth{2}}
                @{\ColumnGap}
                }
                &
                \scriptsize Command slope (microns)
                \\
                \scriptsize $\quad\;\;\;$ Linearity fit of the different actuators &
                \subfigimg[width=\linewidth,pos=ul,font=\fontfig{\subfigColor}]{}{0.0}{\FigTwoBar}
                \\
                \subfigimg[width=\linewidth,pos=ul,font=\fontfig{black}]{$\quad\quad\quad\;\;$(a)}{0.0}{\FigOneOne} &
                \subfigimg[width=\linewidth,pos=ul,font=\fontfig{\subfigColor}]{$\;$(b)}{0.0}{\FigTwoOne}
        \end{tabular}

        \caption{\label{fig:DM_BMC_linearity} Analysis of the amplitude linearity of the actuators of the BMC DM. See caption of \fig{fig:DM_linearity}. Green square: area on which all the fits are performed.}    
\end{figure}

The linearity of the actuator amplitude with the command is shown in \fig{fig:DM_BMC_linearity}. As already noticed by Delacroix \etal\cite{Delacroix:15}, the linearity range of the DM is limited. As a consequence, the linearity slopes are fitted only for the command $c\in\Brack{-\SI{30}{\percent},+\SI{60}{\percent}}$ and no offset is fitted. From \subfig{fig:DM_BMC_linearity}{a}, it appears that between \SI{-40}{\percent} and \SI{70}{\percent} the DM has a good linear response, diverging less that \SI{4}{\percent} from linearity. Beyond this range, it saturates as expected at \SI{100}{\percent}. On the negative values nonetheless, the a strong loss of linearity appears earlier at \SI{-50}{\percent} with a saturation value around \SI{-60}{\percent}. \subfig{fig:DM_BMC_linearity}{b} shows that the command slopes of the actuators are homogeneous across the DM pupil with an average value of \SI{2.5}{\micro\meter}, with a negligible dispersion. The slightly lower values on the side (blue) imply that pushing all the actuators with the same command will produce a small defocus aberration.

The linearity issues mentioned in \figs{fig:DM_BMC_profile}{fig:DM_BMC_linearity} do not represent a big concern. Indeed, this BMC DM is not intended to be used for an adaptive loop where the linearity of the DM response is essential to model its closed-loop behaviour. In the present work, it is used in a static environment to correct for the static aberrations of the bench to reach diffraction limited performances. These non-linearities do not prevent the optimisation algorithms to converge towards the correct membrane shape.

\begin{figure}[ht!]
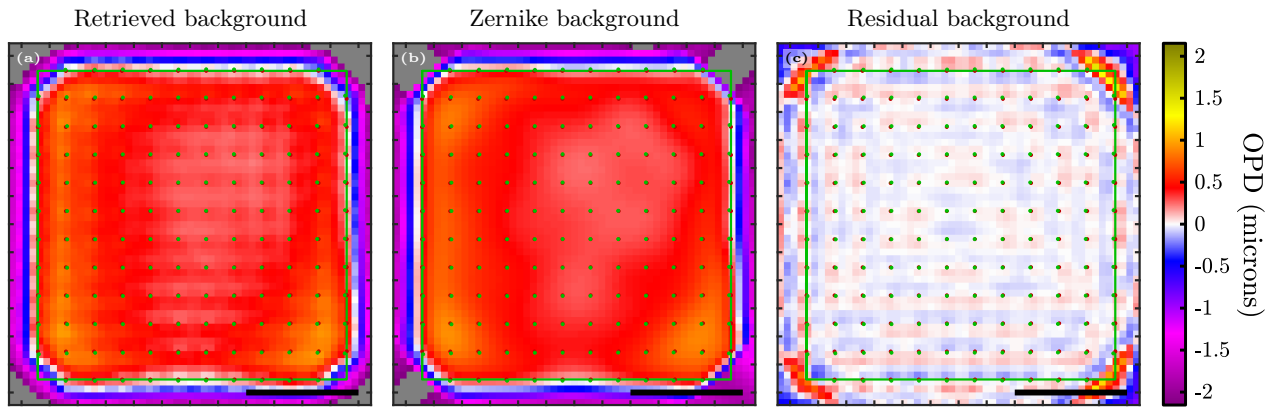
 
        \centering
        
        \newcommand{\LineRatio}{0.95}
        
        \newcommand{\PathFig}{figures_DM_BMC_charac_}

        \newcommand{\Suffix}{_C_p020}
        
        \newcommand{\ExtFig}{.pdf}
        
        \newcommand{\FigOneOne}{\PathFig BG_dat\Suffix \ExtFig}
        \newcommand{\FigTwoOne}{\PathFig BG_fit\Suffix \ExtFig}
        \newcommand{\FigThreeOne}{\PathFig BG_res\Suffix \ExtFig}
        \newcommand{\FigFourOne}{\PathFig Map_BG_bar\ExtFig}

        \newcommand{\subfigColor}{white}        
        
        \sbox1{\includegraphics{\FigOneOne}}               
        \sbox2{\includegraphics{\FigTwoOne}}               
        \sbox3{\includegraphics{\FigThreeOne}}               
        \sbox4{\includegraphics{\FigFourOne}}               
        \newcommand{\ColumnWidth}[1]
                {\dimexpr \LineRatio \linewidth * \AspectRatio{#1} / (\AspectRatio{1} + \AspectRatio{2} + \AspectRatio{3} + \AspectRatio{4}) \relax
                }
        \newcommand{\ColumnGap}{\hspace {\dimexpr \linewidth /5 - \LineRatio\linewidth /5 }}

        \begin{tabular}{
                @{\ColumnGap}
                C{\ColumnWidth{1}}
                @{\ColumnGap}
                C{\ColumnWidth{2}}
                @{\ColumnGap}
                C{\ColumnWidth{3}}
                @{\ColumnGap}
                C{\ColumnWidth{4}}
                @{\ColumnGap}
                }
                \footnotesize Retrieved background &
                \footnotesize Zernike background &
                \footnotesize Residual background &
                \\
                \subfigimg[width=\linewidth,pos=ul,font=\fontfig{\subfigColor}]{$\;$(a)}{0.0}{\FigOneOne} &
                \subfigimg[width=\linewidth,pos=ul,font=\fontfig{\subfigColor}]{$\;$(b)}{0.0}{\FigTwoOne} &
                \subfigimg[width=\linewidth,pos=ul,font=\fontfig{black}]{$\;$(c)}{0.5}{\FigThreeOne} &
                \subfigimg[width=\linewidth,pos=ul,font=\fontfig{\subfigColor}]{}{0.0}{\FigFourOne}
        \end{tabular}

        \caption{\label{fig:DM_BMC_background} Analysis of the OPD background of the BMC DM retrieved from the influence function measurements at the amplitude command $+\SI{20}{\percent}$. See caption of \fig{fig:DM_background}.}
\end{figure}

\begin{figure}[ht!]
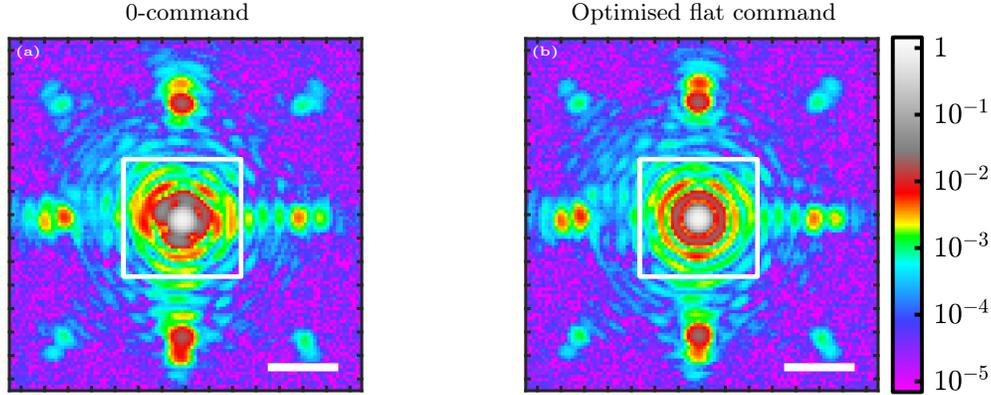
 
        \centering
        
        \newcommand{\LineRatio}{0.65}
        
        \newcommand{\PathFig}{figures_PSF_EvWaCo_}
        \newcommand{\ExtFig}{.pdf}

        \newcommand{\FigOneOne}{\PathFig PSF_0\ExtFig}
        \newcommand{\FigTwoOne}{\PathFig PSF_NCPA\ExtFig}
        \newcommand{\FigThreeOne}{\PathFig Colorbar\ExtFig}

        \newcommand{\subfigColor}{white}        
        
        \sbox1{\includegraphics{\FigOneOne}}               
        \sbox2{\includegraphics{\FigTwoOne}}               
        \sbox3{\includegraphics{\FigThreeOne}}               
        \newcommand{\ColumnWidth}[1]
                {\dimexpr \LineRatio \linewidth * \AspectRatio{#1} / (\AspectRatio{1} + \AspectRatio{2} + \AspectRatio{3}) \relax
                }
        \newcommand{\ColumnGap}{\hspace {\dimexpr \linewidth /3 - \LineRatio\linewidth /3 }}

        \begin{tabular}{
                @{\ColumnGap}
                C{\ColumnWidth{1}}
                @{\ColumnGap}
                C{\ColumnWidth{2}}
                @{}
                C{\ColumnWidth{3}}
                @{\ColumnGap}
                }
                \footnotesize $0$-command &
                \footnotesize Optimised flat command &
                \\
                \subfigimg[width=\linewidth,pos=ul,font=\fontfig{\subfigColor}]{$\;$(a)}{0.0}{\FigOneOne} &
                \subfigimg[width=\linewidth,pos=ul,font=\fontfig{\subfigColor}]{$\;$(b)}{0.0}{\FigTwoOne} &
                \subfigimg[width=\linewidth,pos=ul,font=\fontfig{black}]{}{0.5}{\FigThreeOne}
        \end{tabular}

        \caption{\label{fig:PSF_EvWaCo} Regular PSF (without coronagraph) on the EvWaCo testbed using the BMC DM. White squares: AO corrected area. Scale bar: $\SI{6}{\lambda/D}$. (a)~Setting all the actuators at a command of $+\SI{0}{\percent}$ (0.5 of the BMC SDK). (b)~Preliminary results with an optimised flat command.}
\end{figure}

As already mentioned, along with the influence function measurement, it is possible to retrieve the background phase of the EvWaCo bench, shown in \subfig{fig:DM_BMC_background}{a}. By fitting 130 Zernike modes in this phase map, it is possible to split its low spatial frequency content, in \subfig{fig:DM_BMC_background}{b} from the small scale aberrations, in \subfig{fig:DM_BMC_background}{c}. The map of \subfig{fig:DM_BMC_background}{b} shows that contrary to the ALPAO DM192, the BMC DM does not present major large scale aberrations, behaving almost like a flat mirror. This was observed during its alignment where the first diffraction rings of the PSF could be seen even before optimising the flatness of its membrane\footnote{Contrary to the ALPAO DM for which the 0-command produces a speckle field.}, as shown in \subfig{fig:PSF_EvWaCo}{a}. Nonetheless, \subfig{fig:DM_BMC_background}{c} shows that at the scale of the actuator pitch, the membrane presents a strong periodic feature with a peak-to-valley amplitude of \SI{\pm 75}{\nano\meter}. Having a period equalling the actuator pitch, this pattern cannot be corrected by the actuators, degrading the performances of the DM. This periodic pattern behaves as a grating and is at the origin of the secondary diffraction spots observed in \subfig{fig:PSF_EvWaCo}{b}. This effect is more marked along the vertical direction, explaining with the secondary spots along $y$ in \fig{fig:PSF_EvWaCo} are more intense than along the $x$-direction.

\figfull{fig:PSF_EvWaCo} presents very preliminary results obtained on the EvWaCo testbed on a regular PSF (without coronagraph mask) using the BMC DM in a pupil plane. They are 10 actuators across the pupil. The optimal command to apply on the DM is obtained by maximising the PSF variance.

\acknowledgments 
 
The authors acknowledge the supports from Chulalongkorn University’s CUniverse (CUAASC) grant and from the Program Management Unit for Human Resources \& Institutional Development, Research and Innovation, NXPO (grant number B16F630069).

\bibliography{AO_bench} 
\bibliographystyle{spiebib} 

\end{document}